\documentclass[twocolumn,twocolappendix]{aastex631}

\usepackage{amsmath}
\usepackage{bm}
\usepackage{hyperref}
\usepackage{xcolor}
\usepackage{soul}
\usepackage{graphicx}

\definecolor{light-gray}{gray}{0.8}
\newcommand{\V}[1]{\mathbf{#1}}

\newcommand{\approptoinn}[2]{\mathrel{\vcenter{
  \offinterlineskip\halign{\hfil$##$\cr
    #1\propto\cr\noalign{\kern2pt}#1\sim\cr\noalign{\kern-2pt}}}}}

\newcommand{\appropto}{\mathpalette\approptoinn\relax}

\graphicspath{{./}{new_figs/}}

\submitjournal{ApJ}

\shorttitle{Radiative Magnetic Reconnection in 3D}
\shortauthors{Chernoglazov, A. et al.}

\begin{document}

\title{High-Energy Radiation and Ion Acceleration in Three-dimensional Relativistic Magnetic Reconnection with Strong Synchrotron Cooling}

\correspondingauthor{Alexander Chernoglazov}
\email{achernog@umd.edu}

\author[0000-0001-5121-1594]{Alexander Chernoglazov}
\affiliation{Department of Physics, University of Maryland, College Park, MD 20742, USA}
\affiliation{Institute for Research in Electronics and Applied Physics, University of Maryland, College Park, MD 20742, USA}

\author[0000-0001-8939-6862]{Hayk Hakobyan}
\affiliation{Computational Sciences Department, Princeton Plasma Physics Laboratory (PPPL), Princeton, NJ 08540, USA}
\affiliation{Physics Department \& Columbia Astrophysics Laboratory, Columbia University, New York, NY 10027, USA}

\author[0000-0001-7801-0362]{Alexander Philippov}
\affiliation{Department of Physics, University of Maryland, College Park, MD 20742, USA}
\affiliation{Institute for Research in Electronics and Applied Physics, University of Maryland, College Park, MD 20742, USA}

\begin{abstract}

We present the results of 3D particle-in-cell (PIC) simulations that explore relativistic magnetic reconnection in pair plasma with strong synchrotron cooling and a small mass fraction of non-radiating ions. Our results demonstrate that the structure of the current sheet is highly sensitive to the dynamic efficiency of radiative cooling. Specifically, stronger cooling leads to more significant compression of the plasma and magnetic field within the plasmoids. We demonstrate that ions can be efficiently accelerated to energies exceeding the plasma magnetization parameter, $\gg\sigma$, and form a hard power-law energy distribution, $f_i\propto \gamma^{-1}$. This conclusion implies a highly efficient proton acceleration in the magnetospheres of young pulsars. Conversely, the energies of pairs are limited to either $\sigma$ in the strong cooling regime or the radiation burnoff limit, $\gamma_{\rm syn}$, when cooling is weak. We find that the high-energy radiation from pairs above the synchrotron burnoff limit, $\varepsilon_c \approx 16$ MeV, is only efficiently produced in the strong cooling regime, $\gamma_{\rm syn} < \sigma$. In this regime, we find that the spectral cutoff scales as $\varepsilon_{\rm cut}\approx \varepsilon_c (\sigma/\gamma_{\rm syn})$, and the highest energy photons are beamed along the direction of the upstream magnetic field, consistent with the phenomenological models of gamma-ray emission from young pulsars. Furthermore, our results place constraints on the reconnection-driven models of gamma-ray flares in the Crab Nebula. 

\end{abstract}

\keywords{magnetic reconnection; gamma rays: general; radiation mechanisms: non-thermal; (stars:) pulsars: general; black hole physics}

\section{Introduction} 
\label{sec:int}

Over the past two decades, the physics of magnetic reconnection in the magnetospheres of compact astrophysical objects, such as neutron stars and black holes, has been extensively studied both theoretically and numerically. In particular, the key role of reconnection in the process of magnetic energy dissipation and particle acceleration has been widely recognized and accepted \citep{Lyubarsky1996, Lyubarsky2005, Uzdensky2010, Sironi2014, Uzdensky2014, Uzdensky2016, Uzdensky2016b, Sironi2016, Werner2017}. In some systems, the presence of large-scale reconnection sites has been demonstrated by global simulations, such as pulsar \citep[e.g.,][]{Philippov2014, Chen2014} and black hole magnetospheres \citep[e.g.,][]{Parfrey2019, Bransgrove2021, Ripperda2022}. In other systems, such as the coronae of X-ray binaries or jets of active galactic nuclei (AGNs), the formation of intermittent current layers associated with either turbulent energy cascade or MHD instabilities \citep[e.g., kink instability,][]{Begelman1998, Tchekhovskoy2016, Alves2018, Davelaar2020, Ortuno-Macias2022} has been theorized \citep{Beloborodov2017} or directly demonstrated from intermediate scale simulations \citep{Servidio2011, Zhdankin2013, Zhdankin2017, Zhdankin2018, Comisso2018, Davelaar2020, Chernoglazov2021, Sironi2021}. Relativistic magnetic reconnection has thus established itself as one of the most important plasma-physical processes powering energy dissipation and non-thermal emission in various contexts of high-energy astrophysics.

Strong radiative cooling, both due to synchrotron emission and inverse-Compton scatterings, has implications for particle acceleration and the reconnection process itself \citep[see, e.g.,][]{Uzdensky2016b}. The dynamics of the radiative reconnection has so far been mostly studied for 2D current sheets \citep{Jaroschek2009, Uzdensky2011a, Cerutti2012a, Cerutti2012b, Beloborodov2017, Werner2019, Sironi2020, Mehlhaff2020, Sridhar2021}. For the case of strong cooling (either synchrotron or inverse-Compton), radiative losses lead to the decrease of the plasma temperature, resulting in a strong compression of plasmoids \citep[e.g.,][]{Schoeffler2019, Hakobyan2022b}. Radiative cooling also substantially limits non-thermal particle acceleration. However, in the case of strong synchrotron cooling, the initial acceleration (``injection'') of particles takes place in or close to the X-points, where the radiative cooling is negligible \citep{Cerutti2014, Kagan2016, Hakobyan2019}, allowing the particles to accelerate above the synchrotron ``burn-off'' limit and emit $\gamma$-ray synchrotron photons \citep{Uzdensky2011, Cerutti2012b, Kagan2016b}. 

The case of 3D relativistic current sheets is substantially less explored \citep{Zenitani2008, Guo2014, Sironi2014, Werner2017, Zhang2021, Werner2021, Guo2021, Schoeffler2023}. It has been shown that there are competing instabilities and acceleration mechanisms specific to 3D. For example, \cite{Zenitani2008}, and \cite{Cerutti2014} demonstrated that in the linear stage, the relativistic drift-kink instability grows faster than the plasmoid instability. Later, \cite{Sironi2014}, and \cite{Guo2014} showed that the plasmoid instability still dominates in the non-linear regime, leading to the plasmoid-dominated fast reconnection stage in 3D current sheets and efficient particle acceleration. The 3D plasmoids (or flux tubes) resulting from the plasmoid instability are subject to MHD-kink instability, which ultimately leads to turbulence, the nature and properties of which are largely unexplored \citep{Huang2016, Guo2021}. The mechanisms for particle acceleration are also substantially different between 2D and 3D: while particle ``injection,'' i.e. acceleration to Lorentz factors comparable to the magnetization parameter, $\sigma$, proceeds similarly \citep[e.g.,][]{Sironi2014, Werner2017, Sironi2022}, the acceleration to high energies in 3D is significantly more efficient. In particular, \cite{Dahlin2015}, and \cite{Zhang2021, Zhang2023} have shown that in 3D, particles are capable of escaping the flux tubes and can be directly accelerated by the reconnection electric field. These studies were mainly focused on the non-radiative regime, and it remains to be seen how radiative cooling affects 3D current sheets \citep{Cerutti2014, Schoeffler2023}.

Another important aspect of the problem is the composition of current sheets. While most studies of relativistic reconnection focus on pair plasmas, in many scenarios, such as the magnetospheres of pulsars \citep[e.g.,][see Sec. \ref{subsec:pulsar}]{Guepin2020} and supermassive black holes, we expect a mixture of ions (primarily protons) to be present. Importantly, ions are not susceptible to radiative cooling and have the potential to be accelerated more efficiently by reconnection. Thus far this question, in the context of radiative relativistic reconnection, has not been addressed systematically.

In this paper, we present a set of large 3D PIC simulations of reconnecting current sheets populated by the electron-positron plasma with a small mixture of ions. In our simulations, pairs are subject to dynamically important synchrotron cooling, while ions are unaffected. The goal of our study is to understand the emerging distribution functions of the accelerated ions and pairs, their acceleration mechanisms, and the properties of high-energy radiation. We begin in Sec.~\ref{sec:method} with a description of the numerical setup we employ. We then describe the structure of radiatively-cooled 3D current sheets and show that those with the strong cooling form more compressed flux tubes (Sec.~\ref{sec:csstructure}). In Sec.~\ref{sec:Acceleration} we present our main findings on particle acceleration. In particular, we show that ions are very efficiently accelerated in current sheets with strong radiative cooling, while the Lorentz factors of pairs are limited by the upstream magnetization parameter by the synchrotron losses. In Sec.~\ref{sec:radiation} we discuss the radiation spectra and the beaming of the emission, for varying strength of the synchrotron cooling. We find that 3D current sheets with efficient cooling are capable of emitting high-energy photons above the synchrotron burn-off limit, which are preferentially beamed along the upstream magnetic field. In contrast, current sheets with weak cooling emit photons with energies up to the burn-off limit in the direction perpendicular to the upstream field direction. We summarize our main findings and discuss their implications for magnetospheres of pulsars and black holes in Sec.~\ref{sec:Discussion}.

\begin{table}[b]
\begin{center}
\caption{\label{tab:sim}Summary of simulation parameters.}
\begin{tabular}{ccccc}
\hline \hline
\textrm{Simulation}&
\textrm{$L_x\times L_y\times L_z$}&
\textrm{$d_e/\Delta x$}&
\textrm{$c\Delta t/\Delta x$}&
\textrm{$\gamma_{\rm syn}/\sigma$}\\
\hline
 \texttt{3dx2kCool02} & $2000^2\times 1000$ & 3 & 0.225 & 0.2 \\
 \texttt{3dx2kCool5} & $2000^2\times 1000$ & 3 & 0.225 & 5.0 \\
 \texttt{3dx3kCool02} & $3000^2\times 1000$ & 3 & 0.225 & 0.2 \\
 \texttt{3dx3kCool05} & $3000^2\times 1000$ & 3 & 0.225 & 0.5 \\
 \texttt{3dx3kCool1}  & $3000^2\times 1000$ & 3 & 0.225 & 1.0 \\
 \texttt{3dx3kCool5} & $3000^2\times 1600$ & 3 & 0.225 & 5.0 \\
\texttt{3dx3kCoolInf} & $3000^2\times 1600$ & 3 & 0.225 & $\infty$ \\
 \hline
 \texttt{1dx1kCool02} & $1000^2\times 1000$ & 1 & 0.225 & 0.2 \\
 \texttt{1dx2kCool02} & $2000^2\times 1000$ & 1 & 0.225 & 0.2 \\
 \texttt{1dx1kCool01} & $1000^2\times 1000$ & 1 & 0.1125 & 0.1 \\
 \hline \hline      
\end{tabular}
\end{center} 
\end{table}

\section{Numerical Methods}
\label{sec:method}
For our simulations, we use the \texttt{Tristan v2} multi-species radiative PIC code \citep{tristan_v2}.

\subsection{Initial configuration and boundary conditions}
\label{subsec:ic}
The simulation domain is initialized with a single current sheet in Harris equilibrium. The magnetic field profile is given by $\bm{B} = B_{\rm up} \hat{\bm{x}} \tanh{(z/\delta_{\rm cs})}$, i.e., no guide field is present, where $x$ and $z$ are the coordinates along and transverse to the current sheet, correspondingly, and $\delta_{\rm cs}$ is the initial half-thickness of the current sheet. We initialize uniform thermal upstream pair plasma with a mixture of ions of overall number density, $n_{\rm up}=n_i + n_e + n_{e^+} = 2n_e$, and a low temperature, $T_{\rm up} / (m_e c^2) = 10^{-4}$. In all simulations presented in the main text, the ions constitute a fraction $f_i=0.01$ of the total number of particles\footnote{A simulation with no ions, but otherwise identical to \texttt{1dx1kCool02}, demonstrates similar overall dynamics of the reconnection process.}, and we set the ion-to-electron mass ratio to be $m_i/m_e=1$. This is justified because in the process of reconnection at high pair magnetization, $\sigma=B^2/(4\pi n_e m_e c^2)\gtrsim 10^3$ (for the realistic mass ratio), ions are quickly accelerated to relativistic energies, $\gtrsim \sigma (m_e/m_i)\geq 1$, similarly to pairs \citep[e.g.,][]{Werner2018}. We do not expect a realistic ion-to-electron mass ratio to lead to different results as long as the plasma inertia is dominated by pairs, $n_i m_i/(2 n_e m_e) = f_i\cdot (m_i/m_e) \lesssim 1$.\footnote{This situation applies to the current sheets in pulsar magnetospheres and can be relevant to the pair-dominated coronae of accreting black holes.} These assumptions are validated in the Appendix~\ref{app:concentration}, where we vary both the mass ratio (up to $m_i/m_e=25$) and the fraction of ions, $f_i$. 

The magnetization parameter of the upstream plasma, 
\begin{equation}
\label{eq:sigma}
\sigma\equiv \frac{B_{\rm up}^2/4\pi}{m_e n_{\rm up} c^2\left[\left(1-f_i\right)+f_i\frac{m_i}{m_e}\right]}\approx \frac{B_{\rm up}^2/4\pi}{ m_e n_{\rm up} c^2},
\end{equation} is fixed at $\sigma=50$. The current sheet is initialized as a hot electron-positron-ion plasma of density $n_{\rm cs} = N_{\rm cs} n_{\rm up} \cosh^{-2}{(z/\delta_{\rm cs})}$, and temperature $T_{\rm cs}/(m_e c^2) = 2\sigma / N_{\rm cs}$, where $N_{\rm cs} = 3$ is the maximum overdensity of the layer. Thus, the thermal plasma pressure inside the current sheet balances the pressure of the upstream magnetic field, $B_{\rm up}^2/8\pi$. The initial thickness of the current layer, $\delta_{\rm cs}=7 d_e$, is large enough so the tearing instability is not seeded by the numerical noise. 

Our simulation domain has outflowing boundary conditions in the $\pm \hat{\bm{x}}$-directions, with a thick absorber that gradually damps $\bm{E}$ and $\bm{B}$ fields to zero towards the edges of the box. In the $\pm \hat{\bm{z}}$-direction we implement a moving plasma injector, similar to \cite{Sironi2014, Hakobyan2019, Hakobyan2021}, which provides a fresh supply of magnetized upstream plasma to replenish the outflowing population. In the third direction, $\pm \hat{\bm{y}}$, we impose periodic boundary conditions. The aspect ratio of the domain in the plane of the current sheet is $L_x/L_y=1$, and the size perpendicular to the current sheet is $L_z\sim 0.3...0.5 L_x$, large enough to separate the current sheet and the plasma injector.

To trigger the reconnection, we remove thermal pressure in the central part of the current layer as done in \cite{Sironi2016}. This procedure launches transients leading to the onset of reconnection and moving outward from the center at approximately the speed of light. To minimize the effects of boundary conditions on the statistical steady state, the radiative cooling and the outflow boundaries are switched on when the transient approaches the edge of the simulation box, at $0.5~L_x/c$, where $L_x$ is the extent of the domain in the $x$-direction. In all of our analyses, we exclude the particles from the initial current layer, essentially ignoring the evolution of the sheet in its early transient stage. We consider the current sheet to enter the steady-state regime after about $t\gtrsim 1.5~L_x/c$ from the beginning of the simulation. We run all simulations for $3$ additional light-crossing times, with a total duration of $4.5~L_x/c$. All statistical data is averaged over the final $3$ light-crossing times, while the current sheet is in statistical steady-state.

To study the numerical convergence of our results, we explore resolutions of $d_e/\Delta x = 1...3$, where \mbox{$d_e=\left(m_e c^2/4\pi n_{\rm up} e^2\right)^{1/2}$} is the upstream plasma skin depth, and $\Delta x$ is the cell size ($\Delta x= \Delta y = \Delta z$); the comparison of these two resolutions is presented in Appendix \ref{app:resolution}, where we find only marginal differences in the reconnection dynamics and particle acceleration. The choice of $\sigma$ and $d_e$ determines the characteristic resolution of the Larmor orbits of particles, $r_L^{\rm up}(\gamma)=\gamma m_e c^2/(|e|B_{\rm up})$, accelerated up to $\gamma\sim \sigma$: $r_{L}^{\rm up}(\sigma)/\Delta x = (d_e/\Delta x) \sqrt{\sigma}\approx 7...21$, assuming particles gyrate in the perpendicular magnetic field equal to the upstream value. To better characterize the acceleration mechanisms and minimize the boundary effects, we set our fiducial box sizes to be in the range of $L_x/r_L^{\rm up}(\sigma)\approx 150...300$. Our timestep duration (the CFL number) is fixed at $c\Delta t/\Delta x = 0.225$ for most of our simulations. The summary of all the simulation parameters is given in Table \ref{tab:sim}; a discussion on how the synchrotron radiation is implemented (last column) follows further (see Sec. \ref{subsec:photons}). In our strongest cooling simulation, \texttt{1dx1kCool01}, we employ a halved CFL number, $c \Delta t/\Delta x=0.1125$, compared to our default choice of $0.225$, in order to properly resolve the plasma oscillation period in regions of dense flux tubes, highly compressed due to strong radiative cooling (see Sec.~\ref{sec:csstructure}). The upstream plasma is presented with a total concentration of both electrons and positrons averaging to $1$ particle per cell\footnote{I.e., each cell either contains an electron and a positron injected at the same location to satisfy zero numerical charge or does not contain any simulation particles.}. To test the convergence, we run a simulation identical to \texttt{1dx1kCool02} but with $8$ particles per cell in total and obtain identical results. To mitigate the numerical artifacts from the finite number of particles per cell we employ $8$ digital filter passes on the deposited currents with a $(1/4,1/2,1/4)$ stencil.

To analyze the properties of the current sheet in steady-state, we need to employ a robust method to distinguish the current layer carrying the plasma energized in reconnection from the fresh plasma upstream. The most convenient way is to separately trace particles from the two upstreams ($z\gtrsim 0$ and $z\lesssim 0$ regions), and define the mixing factor, $\mathcal{M}$, as suggested by \cite{Zhang2021},
\begin{equation}
\mathcal{M}=1-2\left|\frac{n_{\uparrow}}{n}-\frac{1}{2} \right|.
\end{equation}
Here $n_{\uparrow}$ is the local number density of particles originating from the $z>0$ region, and $n$ is the total number density. This definition implies $\mathcal{M}=0$ in either of the two upstream regions and $\mathcal{M}\approx 1$ in the mixed region of the current layer. We use the threshold $\mathcal{M} = 0.7$ to separate the upstream from the reconnecting current sheet, however, our results are insensitive to this particular choice, as the upstream-to-downstream transition is quite sharp in terms of the values of $\mathcal{M}$.

\subsection{Radiation}
\label{subsec:photons}

\begin{figure*}[ht]
\includegraphics[width=\textwidth]{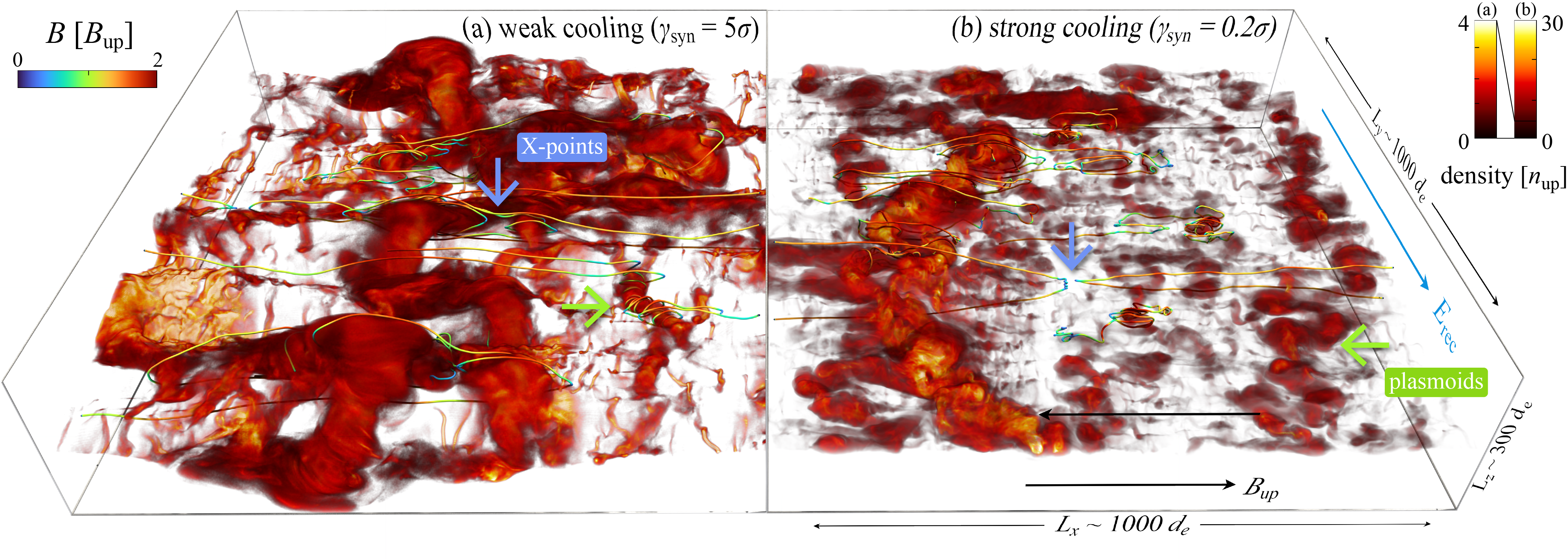}
\caption{Volume rendering of the plasma density in the current sheet with weak (a), and strong (b) cooling at $t\sim 2.2 L_x/c$. Magnetic field lines are also displayed, with the color representing the strength, $|\bm{B}|$. From the field lines, one can easily identify the non-ideal regions -- the X-points (selected ones are shown by blue arrows)-- where the magnetic field vanishes (this corresponds to regions where the field lines are blue). The sizes of plasmoids (selected ones shown by green arrows) are significantly different in these two simulations, with the strong cooling run displaying smaller structures and larger compression (note that the colorbars for the two plots have different scales).}
\label{fig:1_3dsnap}
\end{figure*}

To model the synchrotron cooling, the code imposes a synchro-curvature radiation reaction force in the \cite{Landau1975} form, acting exclusively on the electrons and positrons (ions are unaffected by cooling):
\begin{equation}
\begin{aligned}
{\bm F}_{\rm rad} &= \frac{\sigma_T}{4\pi}({\bm \kappa}_R-\gamma^2 \Tilde{B}_\perp^2 {\bm \beta}),\\
{\bm \kappa}_R &= ({\bm E}+{\bm \beta}\times{\bm B})\times{\bm B}+({\bm \beta}\cdot{\bm E}){\bm E},\\
\Tilde{B}_\perp^2 &= ({\bm E}+{\bm \beta}\times{\bm B})^2-({\bm \beta} \cdot {\bm E})^2.
\label{eqn:cooling}
\end{aligned}
\end{equation}
Here, ${\bm \beta}={\bm v}/c$ is normalized 3-velocity, and \mbox{$\gamma=(1-\beta^2)^{-1/2}$} is the Lorentz factor\footnote{{The total radiated power, $P_{\rm sync}\propto\gamma^2 \Tilde{B}_\perp^2$, corresponds to a square of the electric field in the rest frame of the moving particle, \mbox{${\bm E}'=\gamma({\bm E}+{\bm \beta}\times{\bm B})-(\gamma-1)({\bm \beta} \cdot {\bm E}){\bm \beta}/\beta^2$.}}}. In magnetospheres of compact objects, the radiative drag becomes dynamically important at very high Lorentz factors, $\gamma\gg 1$, depending on the strength of the magnetic field at the location of the reconnecting current sheet. In order to model a similar physical regime in simulations, we rescale the Thomson cross-section $\sigma_T$, using a dimensionless parameter, $\gamma_{\rm syn}$  \citep{Uzdensky2018, Werner2019, Hakobyan2019, Sironi2020}, defined as follows:
\begin{equation}
    |e| E_{\rm rec} \equiv \frac{1}{4\pi} \sigma_T B_{\rm up}^2 \gamma_{\rm syn}^2. 
    \label{eqn:gammaSyn}
\end{equation}
Thus, the value of $\gamma_{\rm syn}$ corresponds to the Lorentz factor of particles for which the synchrotron drag force in the perpendicular background field of $B_{\rm up}$ is equal to the acceleration force $|e| E_{\rm rec}$ (where for magnetic reconnection, we can typically estimate the reconnection-driven electric field $E_{\rm rec}$ as $\approx 0.1~B_{\rm up}$).

The relative importance of radiative cooling for particle acceleration is given by the ratio of $\gamma_{\rm syn}$ and the upstream magnetization parameter, $\sigma$. Since the characteristic energy gain in X-points of the reconnecting sheet corresponds to $\gamma\gtrsim\sigma$ \citep{Sironi2014, Werner2016}, we call the cooling regime ``strong'' if the particles are strongly affected by cooling before their Lorentz factors reach $\sigma$ around the X-point, i.e., if $\gamma_{\rm syn} \lesssim \sigma$. The limit of $\gamma_{\rm syn} > \sigma$ is referred to as the ``weak'' cooling regime.

The second quantity that needs re-scaling is the energy of radiated synchrotron photons, or, equivalently, Planck's constant, $\hbar$. It is easy to see, that the particles with $\gamma\approx \gamma_{\rm syn}$, with $\gamma_{\rm syn}$ being defined as in \eqref{eqn:gammaSyn}, moving in the perpendicular magnetic field of strength $B_{\rm up}$ radiate photons with the characteristic energy
\begin{equation}
    \varepsilon_{c} = \frac{3}{2}\hbar \frac{|e| B_{\rm up}}{m_e c}\gamma_{\rm syn}^2 = \frac{9}{4\alpha_{F}}\frac{E_{\rm rec}}{B_{\rm up}} m_e c^2 \approx 16~\textrm{MeV},
    \label{eqn:burnoff}
\end{equation}
where $\alpha_F\approx 1/137$ is the fine-structure constant \citep[see also][]{Uzdensky2011, Cerutti2012b, Uzdensky2016}. Importantly, this value is insensitive to the strength of the magnetic field and is thus a perfect benchmark to calibrate the photon energies. Thus, in our simulations, a particle with Lorentz factor of $\gamma$ experiencing a magnetic field of $\tilde{B}_\perp$, defined in equation~\eqref{eqn:cooling}, by definition radiates a spectrum of photons which peaks near
\begin{equation}
    \varepsilon_p = 16~\textrm{MeV}~\frac{\tilde{B}_{\perp}}{B_{\rm up}}\left(\frac{\gamma}{\gamma_{\rm syn} }\right)^2. 
    \label{eqn:photonEn}
\end{equation}

\section{Structure of the Current Sheet}
\label{sec:csstructure}

The dynamics of a current sheet, undergoing a 3D magnetic reconnection, is generally similar to that in 2D, but there are also significant differences specific to 3D that must be considered. {The relativistic drift-kink instability is the fastest growing in 3D, and dominates the initial evolution}. As the current sheet evolves, the plasmoid instability in the nonlinear phase becomes dominant \citep[e.g.,][]{Sironi2014}. The dynamical state of the current sheet at late stages is thus dominated by a number of large flux tubes (plasmoids), separated by kinetic-scale X-points. Similar to 2D, the plasmoids are continuously forming and undergoing merging processes. However, contrary to 2D, in 3D they are also being continuously ``dissipated'' by the MHD-kink instability \citep{Werner2021}. 

To visualize the complex 3D structure, Fig.~\ref{fig:1_3dsnap} shows volume renderings of the plasma density in the fully-developed non-linear reconnection stage of 3D current sheets from two different simulations. Fig.~\ref{fig:1_3dsnap}a shows a density snapshot of the simulation with weak cooling  ($\gamma_{\rm syn}/\sigma=5$; \texttt{3dx3kCool5}), while Fig.~\ref{fig:1_3dsnap}b corresponds to the strong cooling case ($\gamma_{\rm syn}/\sigma=0.2$; \texttt{3dx3kCool02}). The small ripples, visible in the perpendicular-to-field direction ($y$-$z$ plane), are caused by the relativistic drift-kink instability. The imprint of the ideal kink instability is visible as the overall twist and deformation of the large flux tubes in both panels of Fig.~\ref{fig:1_3dsnap}. The lines in the figure represent the magnetic field lines and highlight the main features of the current sheet, including the X-points where field lines reconnect crossing the $z=0$ plane, and the plasmoids where magnetic field lines are helically wrapped around high-density structures. The color of the field lines shows the strength of the magnetic field $|\bm{B}|$, particularly demonstrating that it falls to zero in the X-points.

\begin{figure}[t]
\includegraphics[width=\columnwidth]{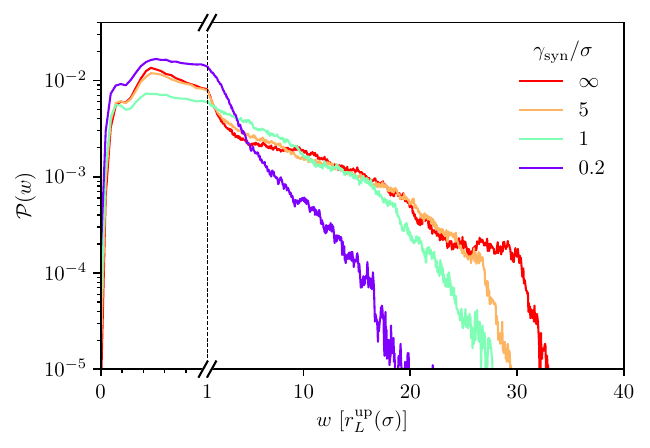}
\caption{Distribution of transverse sizes (along $z$) of the current layer for different cooling regimes. The scale is zoomed in for $w<r_L^{\rm up}(\sigma)$. Stronger cooled current sheets form on average smaller structures due to significant compression. The peak of the distribution, i.e., most of the current sheet, has a thickness below the Larmor radius of particles accelerated up to $\sigma$: $w\sim r_{L}^{\rm up}(\sigma)$.}
\label{fig:2_thickness}
\end{figure}

To systematically quantify the variations of the current sheet structure due to synchrotron cooling, we evaluate the current sheet width, $w$ (along the $z$ axis), for individual points in the horizontal ($x$-$y$) plane using the definition of the mixed region in Sec. \ref{subsec:ic} (this includes both the thin current sheet and plasmoids of various sizes). Fig.~\ref{fig:2_thickness} shows the distribution of widths, $\mathcal{P}(w)$, for a large collection of points for four simulations with varying cooling strength (in the order of increasing cooling strength: \texttt{3dx3kCoolInf}, \texttt{3dx3kCool5}, \texttt{3dx3kCool1}, \texttt{3dx3kCool02}). There are two important conclusions one can draw from this plot. First of all, the peak of the $\mathcal{P}(w)$ is slightly smaller than \mbox{$w_{\rm peak}\sim r_{L}^{\rm up}(\sigma)=d_e \sqrt{\sigma}$}, which corresponds to the Larmor radii of particles with the Lorentz factor close to $\sigma$ in the upstream magnetic field, $B_{\rm up}$. The presence of this peak is associated with the X-points, where particles are accelerated to characteristic Lorentz factors of $\gamma\lesssim\sigma$ (see Sec.~\ref{subsec:IonAcc} and insets of Fig.~\ref{fig:6_ElB}). More importantly, current sheets with stronger cooling display a steeper slope and shorter distribution span for $\mathcal{P}(w)$, indicating that the plasmoids in the current layer become smaller and more concentrated as cooling strength increases (more detailed analysis of this issue was recently performed by \citealt{Schoeffler2023}).

\begin{figure}[t]
\includegraphics[width=\columnwidth]{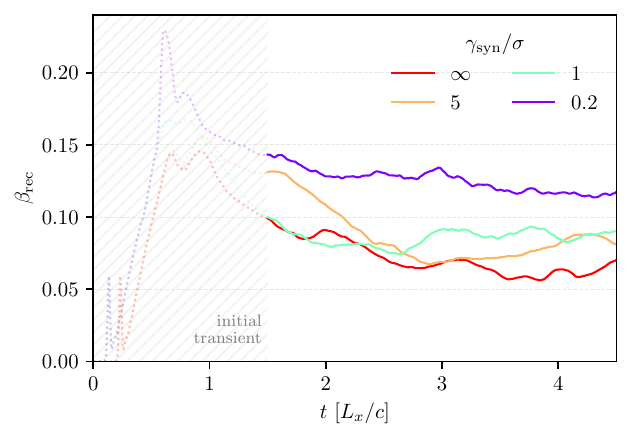}
\caption{
Time evolution of the reconnection rate in simulations with different cooling regimes. Measured rates are of order $\beta_{\rm rec}\approx 0.06...0.14$, and are systematically larger for the case of strongest cooling, $\gamma_{\rm syn}/\sigma=0.2$.
}
\label{fig:3_recrate}
\end{figure}

Variation of the current sheet structure due to synchrotron cooling is reflected in one of the most important macroscopic parameters of the reconnection -- its rate, $\beta_{\rm rec}$. The evolution of the rate, measured as the inflow velocity of the plasma, $\beta_{\rm rec} = \left(\bm{E}\times\bm{B}\right)_z/|\bm{B}|^2$, averaged in the thin layer far upstream parallel to the current sheet, for different cooling regimes is shown in Fig.~\ref{fig:3_recrate}. As mentioned before, we focus our analysis in the time range $t\gtrsim 1.5~L_x/c$, when the simulation reaches a dynamic steady state. We find that the reconnection rate in the weak cooling case ($\gamma_{\rm syn}\gtrsim \sigma$) is $\beta_{\rm rec}\approx 0.06...0.1$ (cf. \citealt{Zhang2021}). Notably, in the simulations with stronger cooling ($\gamma_{\rm syn}\lesssim\sigma$), the reconnection rate is systematically larger, with values ranging around $\beta_{\rm rec}\approx 0.12...0.14$. We associate this correlation with the change in the filling fraction of X-points in the current layer, which for the stronger cooled runs is approximately twice as large, compared to the run without cooling as seen from the peak values in Fig.~\ref{fig:2_thickness}. This conclusion is also consistent with the characteristically smaller plasmoid sizes in the strong cooling regime, as discussed above. 

\begin{figure*}
\includegraphics[width=\textwidth]{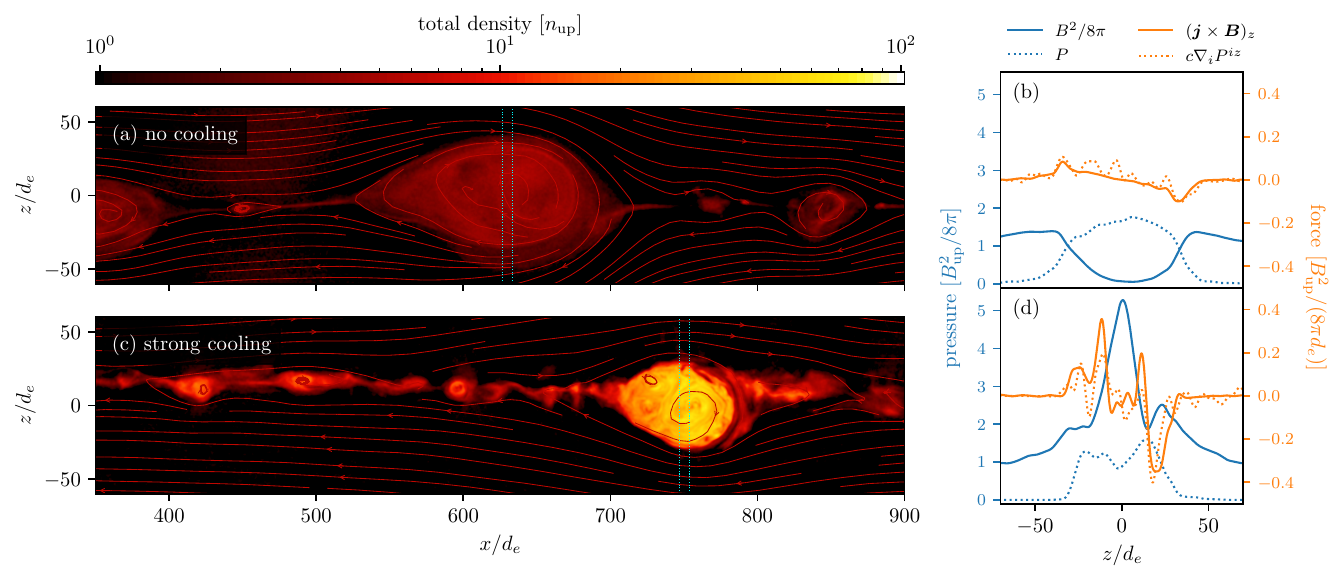}
\caption{
The cross-section of typical plasmoids for uncooled (a, b) and strongly cooled (c, d) current sheets. The left column (a, c) shows density slices and the magnetic field lines. The region highlighted with a cyan-dotted boundary is shown in 1D cuts in (b, d), where we plot the magnetic field, $B^2/8\pi$, and gas pressure, $P$, with the corresponding magnetic tension force, $(\bm{j}\times\bm{B})_z$, and the gradient of the pressure tensor, $c \nabla_{i} P^{i z}$. In the uncooled case, the plasmoids are in equilibrium in terms of magnetic and thermal pressure, with the thermal pressure typically dominating inside the plasmoids. On the other hand, the equilibrium inside the cooled plasmoids is more dynamic, with the magnetic field pressure being larger than the upstream value, and dominating the plasmoid cores.
}
\label{fig:4_pressure}
\end{figure*}

Plasmoids accumulate the hot plasma energized in the X-points. In 2D the energetic particles are well magnetized and confined, resulting in plasmoids being relatively coherent and clearly distinguishable from other relativistic outflows present in the sheet. In 3D the structure of plasmoids is richer and less coherent. Nonetheless, in both cases, their equilibrium is governed by an adiabatic balance between the gradient of thermal pressure of the hot plasma, $P$, and the Lorentz force: $\bm{j}\times\bm{B}=c\nabla P$, with $\bm{j}$ being the total current density. In general, plasma pressure is anisotropic and needs to be computed self-consistently as a moment of the particle distribution function. To compute it, we separate the bulk component of the particle momentum from the thermal one by defining the fluid (Eckart) frame for species $s$, moving with a velocity $U^\mu_s = N^\mu_s/\sqrt{N_{\nu, s} N^\nu_s}$, where $N^\mu_s \equiv \int f_s (u^\mu/u^0) d^3 u$, and $f_s$ and $u^\mu$ are the distribution and the four-velocity of particle species $s$. The plasma pressure tensor for species $s$, $P_s^{\mu\nu}$, is then computed as the projection of the stress-energy tensor: $P^{\mu \nu}_s = \Delta^\mu_\alpha \Delta^\nu_\beta T^{\alpha \beta}_s$, where $T^{\mu \nu}_s \equiv \int (p^\mu p^\nu/p^0) f_s d^3p$, $\Delta^{\mu \nu}_s \equiv \eta^{\mu \nu}+U^\mu_s U^\nu_s$ is the projection operator, and $\eta^{\mu \nu}$ is the Minkowski tensor ($p^\mu=m_s u^\mu$ is the four-momentum of the particle of species $s$). We find that the non-diagonal elements of the pressure tensor compared to diagonal elements are typically at the level of $5\%$ inside plasmoids in the weak-to-no cooling regime, with an increase to non-negligible $\approx 20\%$ for the strongest cooling regime, $\gamma_{\rm syn}/\sigma=0.2$. In addition, inside plasmoids, the diagonal component perpendicular to the magnetic field is suppressed by a factor of $\sim5$ in the case of strong cooling.\footnote{The role of the pressure anisotropy due to synchrotron cooling is discussed by \cite{Zhdankin2022}.} Guided by these findings, we consider the gradient of the full pressure tensor to analyze the force balance of flux tubes.

 In typical 2D simulations (with or without cooling), the force balance inside plasmoids is achieved by strong compression evident in stronger magnetic fields and higher densities \citep{Sironi2016}. The presence of strong cooling exacerbates this effect \footnote{This effect is investigated in detail by \citet{Schoeffler2019}, who observed that this compression provides a positive feedback loop: magnetic field amplification by compression inside plasmoids leads to stronger synchrotron cooling, which in turn promotes further compression.}, by removing the pressure support, which results in extremely compressed structures \citep[see, e.g.,][]{Hakobyan2019}. In 3D, the picture is quantitatively different. The force-balance condition is demonstrated in Fig.~\ref{fig:4_pressure}, where in panels (a) and (c) we show the cross-section of the plasma density in $x$-$z$ plane for two simulations: one without cooling (\texttt{3dx3kCoolInf}), and the other with strong cooling (\texttt{3dx3kCool02}). We pick intermediate-size plasmoids in both of these simulations (indicated by the dotted vertical lines in left panels) and in panels (b) and (d) plot the magnetic pressure, $B^2/8\pi$, the scalar plasma pressure, defined as $P = 1/3\sum_s P^i_{s,~i}$, as well as the $z$-components of the two forces: the pressure gradient, $c\nabla_i P^{i z}$, and the Lorentz force, $\left(\bm{j}\times\bm{B}\right)_z$, as functions of $z$. In the no-cooling simulation (Fig.~\ref{fig:4_pressure}a, b), large plasmoids are typically weakly magnetized, while in the strong cooling regime (Fig.~\ref{fig:4_pressure}c, d), the strength of the magnetic field inside the plasmoids is several times larger than the upstream value, $B_{\rm up}$, (compare blue solid lines in Fig.~\ref{fig:4_pressure}b, d). Despite this, even in the strongest cooling regime, the magnetic field compression inside the plasmoids is much smaller than what is observed in 2D \citep{Hakobyan2021, Hakobyan2022b, Schoeffler2023}. Notably, the strongest $B$-field component inside the plasmoid (in the strong cooling regime) is the out-of-plane one, $B_y$. The generation of this out-of-plane magnetic field component, as well as the dissipation of the in-plane $B_{xz}$ components (evident in the no-cooling regime), are both consequences of the MHD kink instability of 3D flux tubes. This instability dissipates the $x$-$z$ component of the magnetic field, converting it to the plasma thermal energy \citep[e.g.,][]{Werner2021}, while simultaneously acting as a dynamo by generating an out-of-plane field component, $B_y$, until the flux tube becomes stable to kinking \citep{Werner2017, Alves2018, Davelaar2020, ZhangQ2021}. Importantly, this interplay is inherent specifically to 3D reconnection and is thus completely overlooked in 2D.

The plasma inside flux tubes both in uncooled and strongly cooled simulations is compressed and heated to a rough equipartition with the upstream magnetic field pressure: $P\approx B_{\rm up}^2/8\pi$. In the latter case, the pressure is mainly delivered by higher plasma density (compare densities in panels (a) and (c) in Fig.~\ref{fig:4_pressure}), since strong synchrotron cooling prevents plasmoids from heating up by effectively radiating away the components of particle momenta perpendicular to the magnetic field. Orange lines in Fig.~\ref{fig:4_pressure}b and d show the two force terms: the gradient of the plasma pressure tensor, and the magnetic force, both plotted against the vertical axis $z$. The balance between these two forces is well satisfied for the flux tube in the uncooled simulation (Fig.~\ref{fig:4_pressure}b), and is reasonable in the strongly cooled simulation (Fig.~\ref{fig:4_pressure}d).

In this section, we largely neglected the presence of ions (insusceptible to synchrotron cooling) in plasmoids. While their contribution to the total pressure was self-consistently included in our calculations, their relative contribution was typically small for the timescales considered, e.g., in Fig.~\ref{fig:4_pressure}, due to their smaller number density. However, at later times ions can accelerate to much higher Lorentz factors than the pairs (see Sec.~\ref{subsec:IonAcc}), at which point their lack in the number density is compensated by the higher energy. When the ion pressure dominates in flux tubes, we expect the structures of plasmoids to be similar to the case of no-cooling (Fig.~\ref{fig:4_pressure}a, b and Appendix \ref{app:concentration}). The applicability of this regime for the current sheets in pulsar magnetospheres is studied in detail in Sec.~\ref{subsec:pulsar}. 

\section{Particle acceleration}\label{sec:Acceleration}

During the non-linear stage of reconnection, cold particles from upstream -- both pairs and ions -- are advected into the current sheet, where they are energized in either the X-points or the relativistic outflows from the X-points \cite[e.g.,][]{French2022}. After this initial ``injection'' stage most of these particles enter into the plasmoids. In 2D, strong compression of the magnetic field inside plasmoids leads to adiabatically slow secondary acceleration of these particles to high energies limited by the system size \citep{Petropoulou2018, Hakobyan2019}. However, in 3D, the finite size of plasmoids in the out-of-plane direction allows the most energetic particles to freely escape from their bounds. Moreover, as described above, compression of the magnetic field inside plasmoids is significantly smaller in 3D. Thus, the same secondary acceleration mechanism as in 2D cannot operate efficiently in 3D. On the other hand, in 3D, another acceleration channel has been observed for uncooled pair plasmas by \cite{Zhang2021, Zhang2023}, with the large-scale reconnection electric field in the upstream, $E_{\rm rec}$, being its main driver. 

\begin{figure}
\includegraphics[width=\columnwidth]{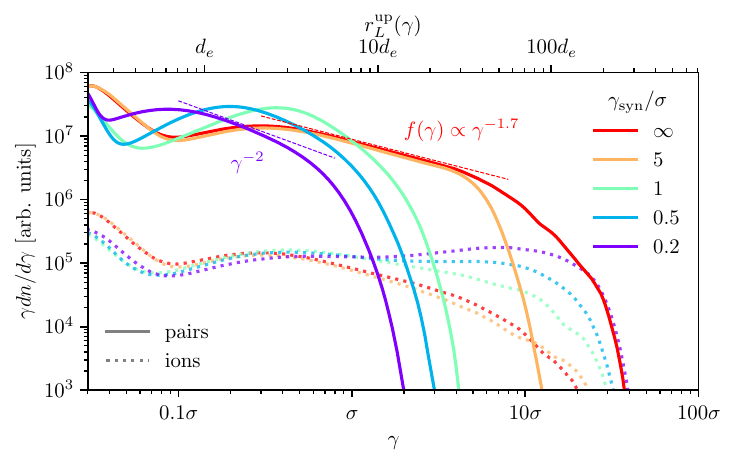}
\caption{Energy distributions of the pairs (solid lines) and the ions (dotted lines) at $t=2.25~ L_x/c$ for different cooling regimes: purple lines correspond to strong cooling \mbox{$\gamma_{\rm syn}/\sigma=0.2$}, and the red ones for the uncooled runs, $\gamma_{\rm syn}/\sigma=\infty$. Dynamically strong cooling effectively limits the maximum energy of pairs to $\sim \sigma$, but at the same time facilitates the formation of harder distribution for the ions. For reference, we also show the corresponding Larmor radii, $r_{L}^{\rm up}$.}
\label{fig:5_energy_distributions}
\end{figure}

The dynamics of plasma in X-points is insensitive to the presence of synchrotron cooling, as the magnetic field vanishes in these regions. However, as soon as the energized pairs leave the X-points, they lose their energy shortly after being exposed to a strong perpendicular magnetic field component. As a result, the secondary acceleration channel in the reconnection upstream is strongly suppressed for pairs. Ions, on the other hand, are unaffected by cooling, and can thus tap the upstream electric field and get accelerated to potentially large Lorentz factors, $\gg\sigma$. This can clearly be seen in Fig.~\ref{fig:5_energy_distributions}, where we plot the
energy spectra of both pairs and ions. For the simulations with strong-to-marginal cooling ($\gamma_{\rm syn}/ \sigma=0.2...1$), the distribution of pairs (solid lines) typically spans up to $\sim\sigma$, with the slope being close to $\gamma^{-2}$ at highest energies. Ions (dashed lines), on the other hand, regardless of the cooling strength, extend to Lorentz factors significantly above $\sigma$, with the cutoff energy itself increasing in time. In the uncooled simulation (red lines), the cutoff in the distribution of pairs is exactly the same as for the ions, suggesting that the underlying secondary acceleration mechanisms are identical. Surprisingly, in simulations with stronger cooling, ions form a much harder power law distribution, with the typical slope close to $\gamma^{-1}$ for the strongest cooling case, and steepening to $\gamma^{-1.7}$ for the uncooled case (this effect is explored in Sec \ref{subsec:IonAcc}).

\begin{figure}
\includegraphics[width=1.0\linewidth]{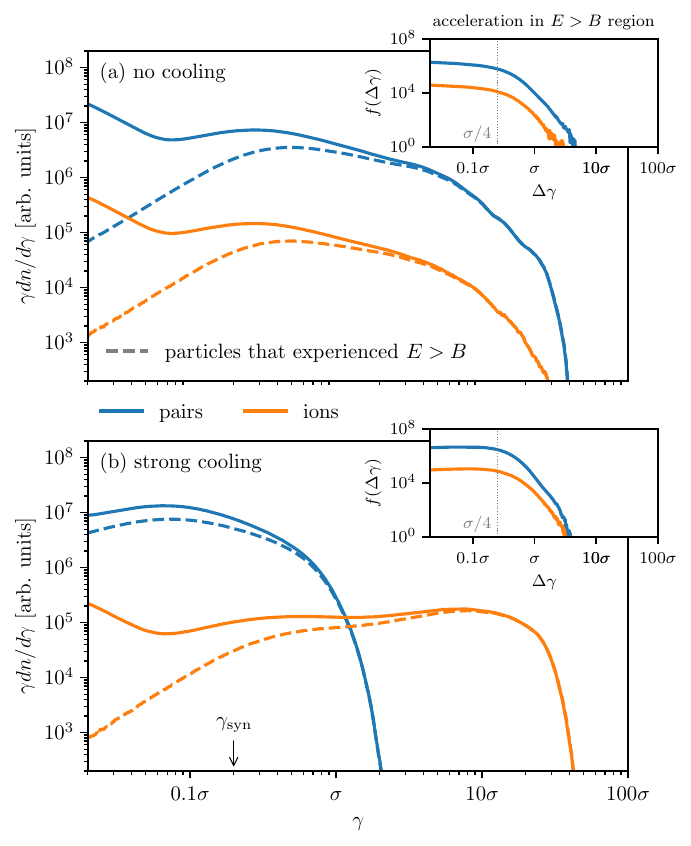}
\caption{Distribution functions of the pairs (blue) and the ions (orange) for both the uncooled (a) and the strongly cooled (b) simulations. Dashed lines show parts of the distributions contributed by the particles having experienced $E>B$ during their lifetime in the sheet. Inset panels show the distribution of energy gains, $\Delta \gamma$, of particles during their contiguous presence in these regions. The non-ideal electric field in the X-points only accelerates particles to Lorentz factors of a fraction of $\sigma$. At the same time, this process is essential to promote the particles (pairs and ions) for further acceleration up to the highest energies: in relativistic outflows along the layer and the large-scale reconnection electric field upstream. As a result, the high energy tail of the spectrum is fully formed by the particles passed through $E>B$.
}
\label{fig:6_ElB}
\end{figure}

We further quantify the contributions of different acceleration channels in Fig.~\ref{fig:6_ElB}, where we plot the energy spectra of both ions and pairs in the uncooled and strongly cooled simulations. To test the dominant acceleration mechanism up to $\gamma\sim\sigma$ (the injection stage), we choose the particles that experienced the non-ideal electric field, $E>B$, during their lifetime (dashed lines in Fig.~\ref{fig:6_ElB}). In the inset panels of Fig.~\ref{fig:6_ElB} we also plot the distribution of energies, $\Delta \gamma$, gained during the passage of $E>B$ regions. Notably, the particles that have passed through an $E>B$ region dominate at energies $\gtrsim\sigma$ \citep[cf.][]{Sironi2022}. However, the net energy gain of particles in X-points in both the uncooled and strongly cooled cases does not exceed a fraction of $\sigma$, and thus cannot account for the high-energy power-law tail of either the pairs or the ions \citep[e.g.,][]{Werner2016, Uzdensky2022}. The main energy gain, thus, occurs due to the acceleration by ideal electric fields. For pairs, as we argue further in Sec.~\ref{subsec:LepAcc}, ``pick-up'' acceleration by outflows from the X-points allows them to gain energies up to $\sigma$ even in the strong cooling case. 

\subsection{Ion acceleration}
\label{subsec:IonAcc}

\begin{figure*}
\includegraphics[width=\textwidth]{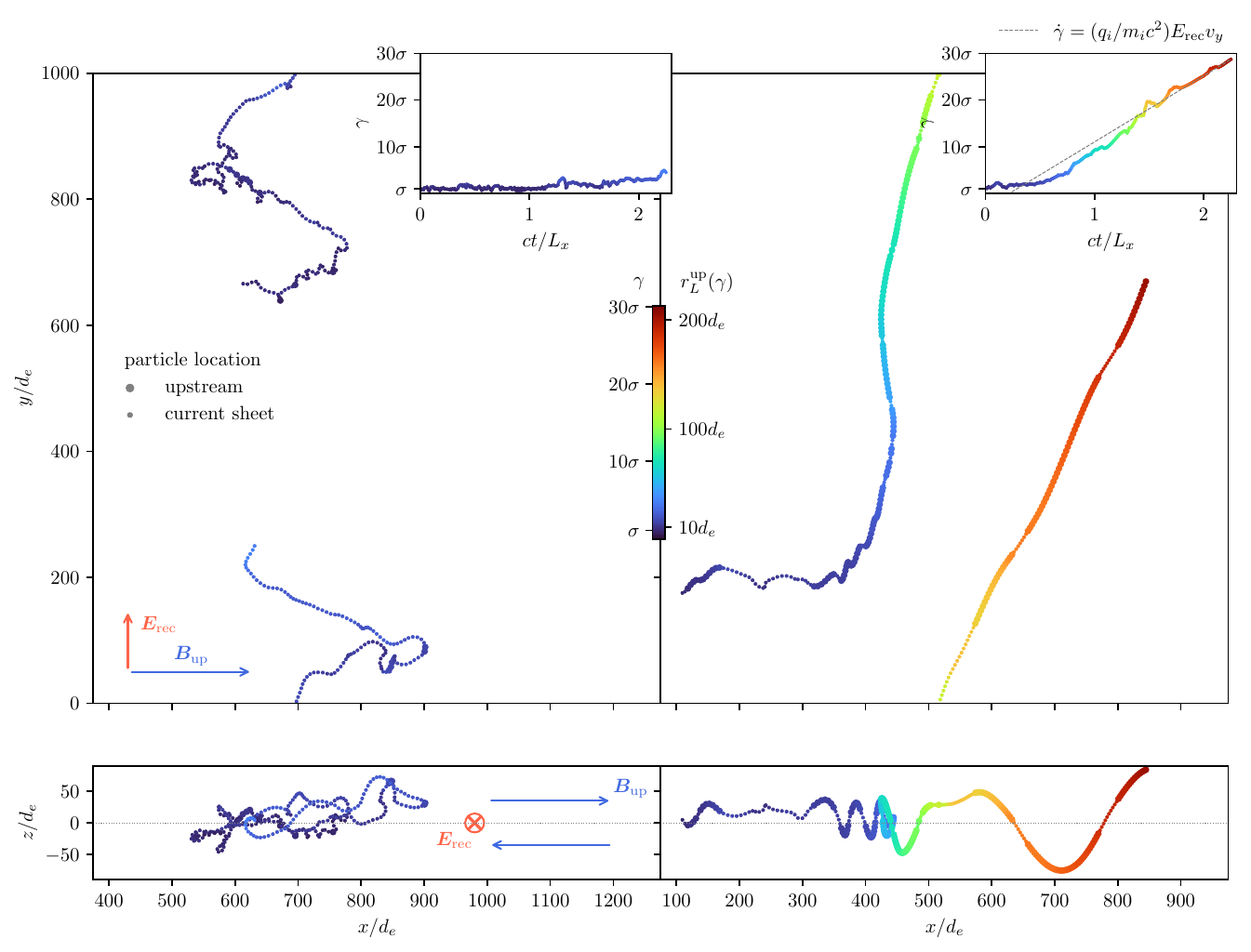}
\caption{Archetypal trajectories (projected to the $x$-$y$ and the $x$-$z$ plane) and the energy evolution of two ions. The color of each dot corresponds to the Lorentz factor of the ion at that particular time, while the size of the dot indicates whether the ion is inside the current sheet (small dots), or upstream (large dots). The ion on the right accelerates to high energies ($\gg \sigma$) by the large-scale electric field, $E_{\rm rec}$, in the region upstream of the layer, while the one on the left gains only a marginal amount of energy ($\sim \sigma$). The more energetic ion on the right primarily moves upstream of the current layer along the reconnection electric field ($y$ direction) and accelerates linearly in time. The slower particle, on the other hand, is trapped inside the current sheet and experiences constant scatterings on local inhomogeneities.}
\label{fig:7_ions}
\end{figure*}

In our simulations, ions are modeled as separate particle species with a mass equal (or comparable) to that of pairs. Additionally, ions do not undergo synchrotron cooling. In Fig.~\ref{fig:7_ions}, we present 2D projections of the ion trajectories obtained from the 3D simulation with strong sycnhrotron cooling, $\gamma_{\rm rad}=0.2\sigma$, \texttt{3dx3kCool02}. Each point on the trajectory is color-coded according to the Lorentz factor of the ion, this information is also displayed on the inset axes. The thickness of the line shows whether the ion is upstream (thick) or inside the current sheet (thin). We initiate the tracking by picking two representative ions with energies $\gamma\sim\sigma$ at $t=2.25~L_x/c$ (corresponding to the same moment shown in Fig.~\ref{fig:5_energy_distributions}, when the spectrum shape is in the steady-state), and follow their evolution until the end of the simulation  $t=4.5 ~L_x/c$. Interestingly, the behavior of the two ions differs significantly. One ion reaches a high Lorentz factor, $\sim 30\sigma$, while the energy of the other ion oscillates around a few $\sigma$. Notably, the ion that gains large energy experiences most of its acceleration while moving outside of the current layer \citep{Zhang2021,Zhang2023}, as evident from the thickness of the line representing the trajectory in Fig.~\ref{fig:7_ions}. These trajectories correspond to bouncing of accelerated ions between the two converging upstream flows\footnote{Note that the classical Speiser orbits correspond to the acceleration inside the current sheet, as opposed to the acceleration upstream (for more discussion, see \citealt{Zhang2021}).}. In contrast, the low-energy ion displays a tangled trajectory due to its entrapment inside a plasmoid or frequent scattering on magnetic field inhomogeneities inside the current sheet. This behavior is likely associated with self-generated turbulence \citep[e.g.,][]{Guo2021}.  

\begin{figure*}
\includegraphics[width=\textwidth]{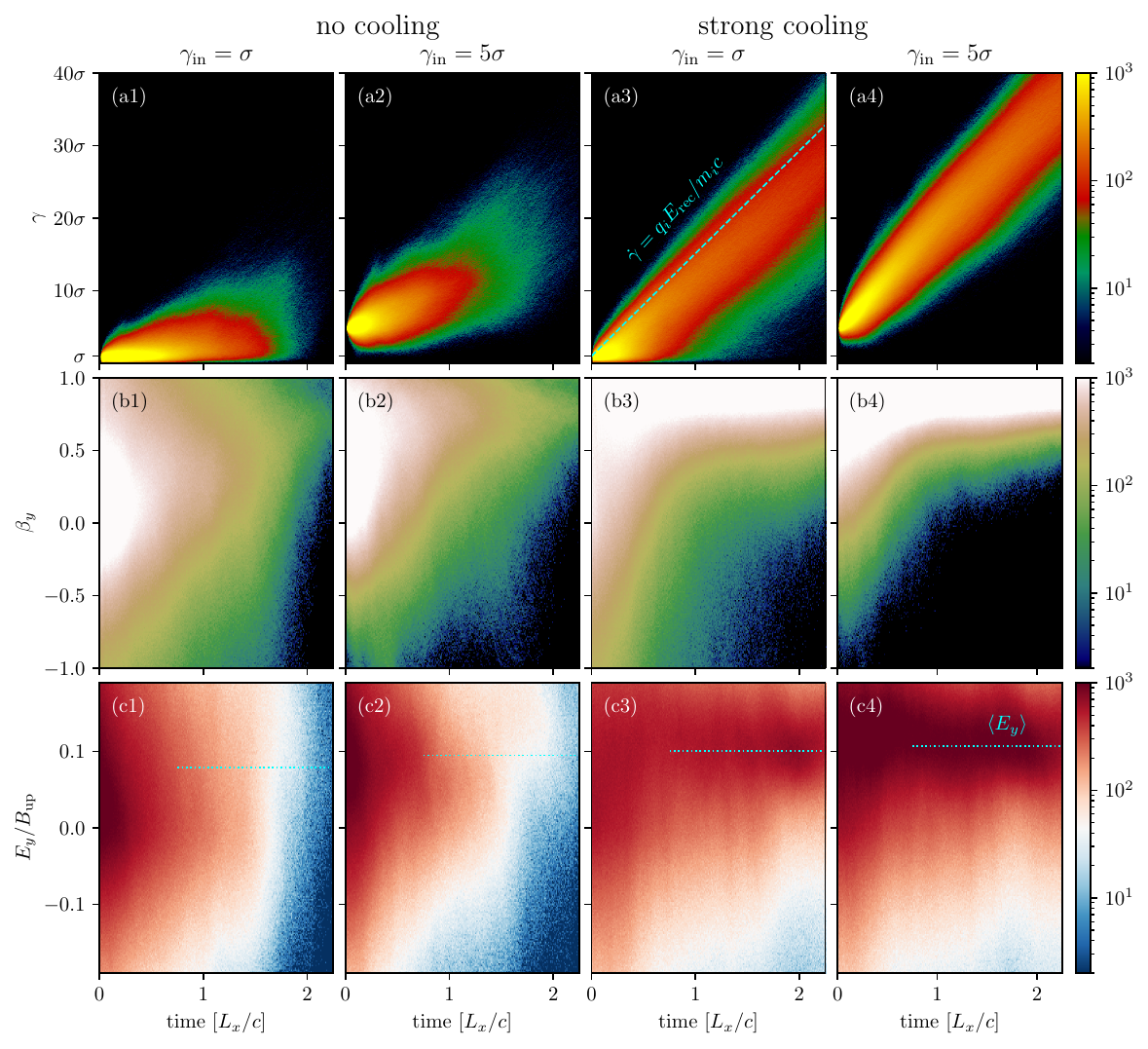}
\caption{Time evolution of a sample of ions that reach energies, correspondingly, $\gamma_{\rm in}=\sigma$ (odd columns), and $\gamma_{\rm in}=5\sigma$ (even columns). Results from two simulations are shown: with no cooling (first two columns), and strong cooling current (last two columns). The upper row (a1...a4) shows the evolution of the Lorentz factor of the ions. The middle row (b1...b4) shows the evolution of the $y$ component of their 3-velocities (along the reconnecting electric field $E_{\rm rec}$). The lower row (c1...c4) shows the value of the reconnection electric field $E_y$ experienced by these ions. In the strong cooling regime, most of the ions successfully accelerate in the reconnection electric field and experience little-to-no scatterings (evident from the narrow spread in their energy evolution, a3-a4). In the uncooled regime, the energy gain is more chaotic as ions constantly scatter inside the current sheet.}
 \label{fig:8_statistics}
\end{figure*}

The acceleration of the more energetic ion takes place linearly in time, with its trajectory showing almost no scatterings. Notably, the preferred direction of motion for this ion is in $y$, while the acceleration rate is almost constant. Both of these facts point towards the large-scale reconnection-driven electric field, $\bm{E}_{\rm rec} \approx \bm{E}_y$, being the dominant acceleration mechanism. The rate of acceleration for these trajectories can be written as
\begin{equation}
\dot{\gamma} = \frac{q_i}{m_i c} \left(\bm{E}_{\rm rec} \cdot \frac{\bm{v}}{c}\right) \Longrightarrow \Delta \gamma \approx \beta_{\rm rec} \beta_y \frac{q_i B_{\rm up}}{m_i c} \Delta t,
    \label{eqn:partAcc}
\end{equation}
where we used $E_{\rm rec} = \beta_{\rm rec} B_{\rm up}$ \citep{Zhang2021}. This slope is overplotted with a dashed line in the inset of the left panel of Fig.~\ref{fig:7_ions}, showing excellent agreement with the acceleration history. Since ions are by construction not susceptible to synchrotron cooling, we find that this acceleration mechanism operates similarly for both the strong cooling and no-cooling regimes.

Despite the ion acceleration mechanism being identical for different cooling regimes, the distribution of ions is much harder in the stronger cooling cases, as highlighted with dashed lines in Fig.~\ref{fig:5_energy_distributions}. Two parameters from Eq.~\eqref{eqn:partAcc} that could potentially contribute to different acceleration rates are the reconnection rate, $\beta_{\rm rec}$,\footnote{As it was already mentioned in Sec.~\ref{sec:csstructure},  the global reconnection rate is larger for the strongest cooling regime, which in turn means that the large-scale electric field, $E_y$, is typically stronger.} and the beaming of ions towards the direction of the reconnection electric field, $\beta_y$. To inspect the relative importance of the two effects, in Fig.~\ref{fig:8_statistics} we plot time-dependent statistical distributions of particle Lorentz factors, $\gamma$ (panels a1...a4), their velocity component along the reconnection electric field in $y$, $\beta_y$ (panels b1...b4), and the electric field strength in $y$ (panels c1...c4). The first two columns represent ions from the uncooled simulation, while the last two -- from the simulation with the strongest cooling, $\gamma_{\rm syn}/\sigma = 0.2$. Since we are primarily interested in the post-injection phase, in the odd-numbered columns we select ions that have reached energies $\gamma_{\rm in}\approx \sigma$ (the core of the distribution), while in the even-numbered ones, we pick ions with $\gamma_{\rm in}\approx 5\sigma$ (the middle of the tail of the distribution). 

If we consider the population of particles with energies $\gamma\approx\sigma$ (consider a slice at $t=0$ in Fig.~\ref{fig:8_statistics}b1, b3), we see that the motion of ions is either not beamed at all in the $y$-direction (the case with no cooling), or has only marginal beaming (the case with strong cooling). By this time, the energy gain of these ions, $\gamma_{\rm in}\approx \sigma$, is dominated by the Fermi acceleration in the current sheet (either via a Fermi kick or the ``pick-up'' mechanism). This prelude is consistent with the fact that the ions at these energies are primarily moving in the plane of the reconnecting magnetic field, $x$-$z$, i.e., perpendicular to the reconnection electric field, $\beta_y\approx 0$, \citep[cf.][]{French2022}. At a later time, a fraction of these particles begin accelerating upstream of the layer along the $y$ direction, i.e., along $\bm{E}_{\rm rec}$; that fraction of ``lucky'' particles ultimately determines the power-law slope in the distribution. As it is evident from the first column of Fig.~\ref{fig:8_statistics}, in the uncooled run only a small fraction of ions follows the $\dot{\gamma}\propto \bm{E}_{\rm rec}\cdot \bm{v}$ linear acceleration path, with the bulk beaming in $y$-direction remaining marginal. This is due to the fact, that the cross-section of typical plasmoids in the simulation with weak-to-no cooling is significantly larger (see Fig.~\ref{fig:2_thickness}), making them more capable of trapping the freely accelerating ions. On the other hand, in the strong cooling case almost all of the particles that reach $\gamma_{\rm in}\approx \sigma$ are ``lucky'' to land on the efficiently accelerating trajectories, beamed in the direction of the electric field in the upstream  (see the third column in Fig.~\ref{fig:8_statistics}; these trajectories are similar to the one shown in the right column of Fig.~\ref{fig:7_ions}). As the ``lucky'' particles become more energized, they are essentially bound to remain on these accelerating trajectories. This can be clearly seen in the second and the fourth panels of Fig.~\ref{fig:8_statistics}, where we show the same tracking statistics, but only for the particles that have already reached $\gamma_{\rm in}\approx 5\sigma$. For this selection, the fraction of ``lucky'' particles is substantially larger. This suggests that even in the uncooled case the ion energization mechanism to $\gamma\gg \sigma$ is the same large-scale acceleration in the upstream, albeit a smaller amount of accelerated particles, free-streaming in the upstream, results in a steeper power-law slope in their distribution.

We observe a general tendency that the acceleration rate for individual ``lucky'' ions is larger for the strongly cooled simulations. This can be clearly seen from the typical strength of the electric field experienced by the ions in Fig.~\ref{fig:8_statistics}c1...c4: for stronger cooling the electric field is systematically larger than in the uncooled case, even if we consider particles beyond $\gamma_{\rm in}\approx 5\sigma$. We associate this difference with the fact that the global reconnection rate, $\beta_{\rm rec}$, is larger for the strong cooling case, as shown in Fig.~\ref{fig:3_recrate}, which in turn means larger $E_y=E_{\rm rec}\approx \beta_{\rm rec}B_{\rm up}$. Another notable feature is the evolution of the distribution of electric field values, $E_y$, felt by ions (especially for $\gamma_{\rm in}\approx \sigma$, panels c1 and c3). At early times, while particles still have relatively low energies, they move primarily in the current sheet, where the electric field is mostly chaotic. When accelerated to large-enough energies, most of these ions are moving primarily in the upstream (cf. Fig.~\ref{fig:7_ions}), where the electric field is coherent and almost exclusively points in the $y$-direction.

\subsection{Pair acceleration}\label{subsec:LepAcc}

Acceleration mechanisms for pairs in the weak-to-no cooling regimes are identical to that for ions: pairs are first accelerated in the X-points and their outflows by non-ideal and ideal electric fields up to Lorentz factors comparable to $\sigma$, after which a fraction of them escape to the upstream where they are linearly accelerated further by the large-scale $E_{\rm rec}$. However, when the cooling is enabled, even if it is dynamically weak, $\gamma_{\rm syn}\gtrsim \sigma$, this linear acceleration cannot energize the pairs to arbitrarily high energies. In this case, the typical value of the effective perpendicular magnetic field, $\tilde{B}$, defined in \eqref{eqn:cooling}, for particles accelerating upstream is close to $B_{\rm up}$. This implies that the upstream acceleration in the weak cooling regime is limited by the burnoff limit, $\gamma \lesssim \gamma_{\rm syn}$. In Fig.~\ref{fig:5_energy_distributions}, this effect can clearly be seen by comparing the uncooled simulation with the one with $\gamma_{\rm syn}=5\sigma$. While in the uncooled case, the distributions of both pairs and ions extend to several tens of $\sigma$ in energy, in the weakly cooled simulation the distribution of pairs essentially cuts off at $\approx \gamma_{\rm syn} = 5\sigma$, with ions being relatively unaffected.

In the strong cooling regime, the upstream acceleration is essentially prohibited, as the strong synchrotron cooling disallows particles to leave the X-points by crossing magnetic field lines. As shown in Fig.~\ref{fig:5_energy_distributions}, the spectrum of accelerated leptons does not extend beyond $\approx \sigma$. In this regime, two acceleration mechanisms are particularly efficient: the X-point  acceleration by the non-ideal electric field, and the ``pick-up'' mechanism. In Fig.~\ref{fig:6_ElB}b, we show that particle energy gain in the $E>B$ zones is not sufficient to extend the power-law to $\gamma\approx\sigma$; this implies that pairs with the highest energies are additionally accelerated by the ``pick-up'' mechanism. 

\begin{figure}[t!]
\includegraphics[width=\columnwidth]{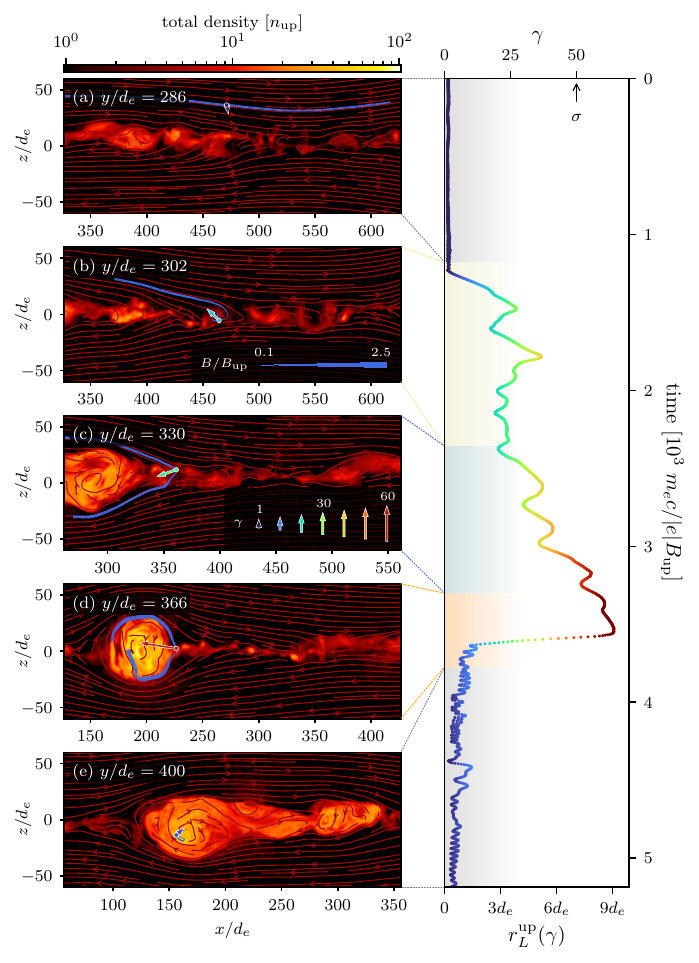}
\caption{Time evolution of a typical accelerated electron in the strongly cooled simulation. The right panel shows the energy evolution of the particle with time. Panels (a)...(e) show the density slice and the magnetic field lines in the $x$-$z$ plane at a particular time of the simulation (corresponding to the right panel). The position and the velocity of the electron are shown with a colored circle and an arrow (color corresponds to its energy at that particular time). The magnetic field line passing through the position of the particle is shown as a blue line, with its thickness corresponding to the strength of the magnetic field. The electron starts cold upstream of the current layer (a) and is accelerated to the Lorentz factor up to $\approx \sigma$ in the current sheet (b), after which it is picked up by the outflow from the X-point and gains energy further (c). Ultimately, the electron experiences an intensive slow-down when it collides with the magnetic field inside the plasmoid (d), where it remains throughout the rest of its lifetime (e).}
\label{fig:9_traj}
\end{figure}

\begin{figure*}[htb]
\centering
\includegraphics[width=0.9\textwidth]{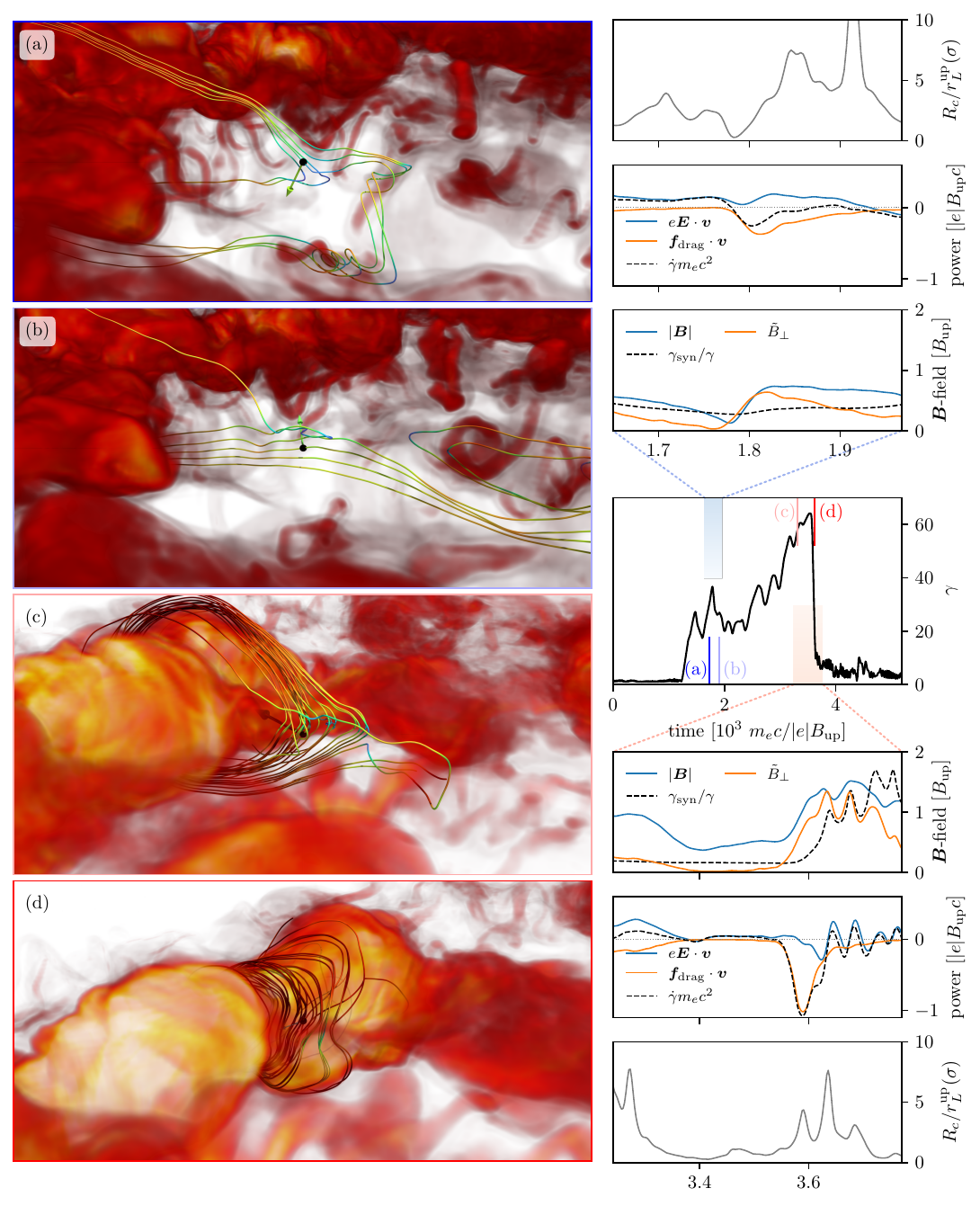}
\caption{Close-up view of the acceleration dynamics of the electron shown in Fig.~\ref{fig:9_traj}. Panels (a...d) show 3D snapshots of the density zoomed in at the position of the particle (shown with a black sphere). We also show the vector of its velocity, and the magnetic field lines close to the location of the particle. Line colors represent the amplitude of the magnetic field, with the colorbar being identical to the one in Fig.~\ref{fig:1_3dsnap}. The middle panel on the right shows the energy evolution of the electron, with the moments of the 3D snapshots indicated by vertical lines. Three extra panels at the top and the bottom of the right column show the time evolution of several important quantities, limited in time to the intervals of the scattering (top), and the catastrophic cooling events (bottom). In particular, we plot the strength of the magnetic field, and the $\tilde{B}_\perp$, experienced by the electron, as well as the critical magnetic field, $B_{\rm up} \gamma_{\rm syn}/\gamma$. We also show the contributions of two forces, the acceleration by the electric field, $e\bm{E}\cdot\bm{v}$, and the cooling, $\bm{f}_{\rm drag}\cdot \bm{v}$, to the energy evolution of the particle, $\dot{\gamma}$. Finally, we show the curvature radius of the magnetic field at the location of the particle.}
\label{fig:10_11_traj_cool}
\end{figure*}

\begin{figure*}[htb]
\includegraphics[width=\textwidth]{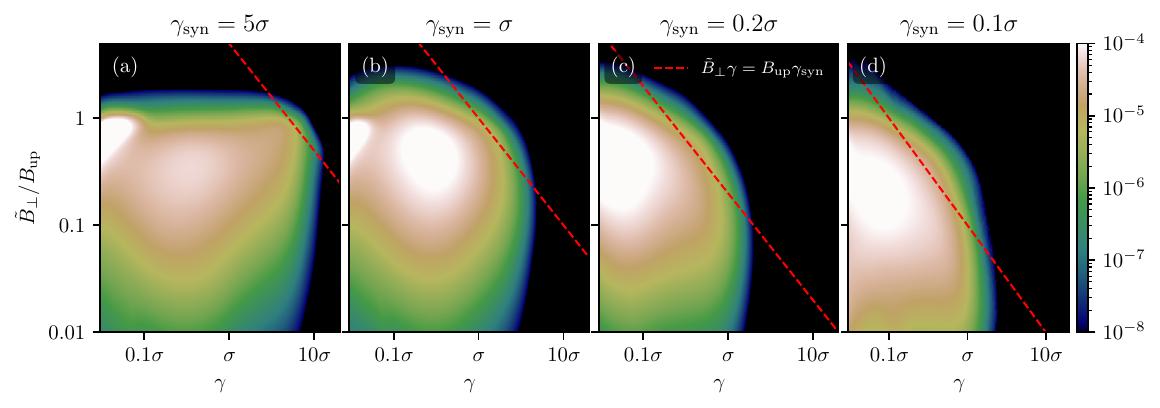}
\caption{The effective perpendicular magnetic field, $\Tilde{B}_\perp$, experienced by the pairs of different energies, $\gamma$. The 2D histograms are shown for all the cooling regimes, from weak, $\gamma_{\rm syn}/\sigma=5$, to strong, $\gamma_{\rm syn}/\sigma=0.1$. The red dashed lines show the estimated critical magnetic field, $\tilde{{B}}_\perp\gamma=B_{\rm up}\gamma_{\rm syn}$. Notably, the acceleration of the highest-energy pairs is inhibited by the synchrotron cooling, and, as a result, most of the pairs do not cross the critical line.}
\label{fig:11_hist}
\end{figure*}

An example of a typical high-energy electron trajectory in the strong cooling regime (\texttt{3dx2kCool02}) is shown in Fig.~\ref{fig:9_traj}. In panels (a)...(e) we plot the $x$-$z$ slices of the plasma density, and the magnetic field lines at different timesteps, with the current positions of the particle indicated with colored circles. The instantaneous four-velocities of the particle are indicated with colored arrows. We also over-plot in blue the magnetic field line at the position of the particle, with the width of the line indicating its strength. On the right of Fig.~\ref{fig:9_traj} we show the time evolution of the Lorentz factor $\gamma$, with the dashed lines indicating the times of the snapshots on the left. In panel (a) the particle is upstream from the layer, moving towards the current sheet with a small Lorentz factor, $\gamma\sim 1$. In panel (b), the particle enters the X-point and is accelerated by the non-ideal electric field, $E>B$, to $\gamma\approx 0.5\sigma$. After this rapid acceleration event, the particle moves along with the relativistic outflow from the X-point constantly crossing the upstream/downstream boundary, which is perturbed vertically due to RDKI (see below). As the particle hits the upstream, it experiences subtle energy losses indicated as oscillations in its energy, $\gamma$. After that, the particle is picked up by a contracting magnetic field line performing a slingshot-like motion. After that, as shown in panel (c), for an extended period of time particle moves along with a rapidly receding field line while being slowly accelerated to $\gamma\gtrsim\sigma$. Once the slingshot-like motion of the field line comes to a halt, the particle is injected into the forming plasmoid, and experiences a catastrophic cooling event, being exposed to a large $\tilde{B}_\perp$ (panel (d); this event is studied in more details later). During its interaction with the plasmoid, the particle radiates most of its momentum and is then trapped inside the plasmoid until the end of the simulation, with a low Lorentz factor, $\gamma\sim \text{few}$, as shown in panel (e).

We further describe the 3D motion of the same accelerated electron in Fig.~\ref{fig:10_11_traj_cool}a...d, where we show 3D volume renderings of the plasma density and the magnetic field lines in the vicinity of the electron, also shown in Fig.~\ref{fig:9_traj}. Panels on the right display the energization history of the particle, as well as the interesting quantities zoomed at two distinct moments of the particle's evolution (roughly corresponding to panels (a)/(b), and (c)/(d) of the same figure). In the upper half of Fig.~\ref{fig:10_11_traj_cool}, we specifically focus on the time when the particle experiences successive scattering events while moving in the RDKI-perturbed current layer,\footnote{A 3D movie of this moment is available via the following link: \url{https://youtu.be/AILs-clOaPY}.} after being pre-energized in the X-point (corresponding to Fig. \ref{fig:9_traj}b). As indicated in the upper right panels, just before the scattering event, the particle crosses the null line, $|\bm{B}|\approx 0$, and approaches the peak of the RDKI perturbation, where both $\tilde{B}_\perp$ and $|\bm{B}|$ grow rapidly (time $t\approx 1.8$ in units indicated on the plot; field strengths at particle position are shown with blue and orange lines). $\tilde{B}_\perp$ reaches the critical value, $\approx B_{\rm up}\gamma_{\rm syn}/\gamma$ (see below), after which deceleration occurs due to the radiative drag force, $\bm{f}_{\rm drag}$ (shown with an orange line), with the particle being slowed down until the field drops below critical (time $t\approx 1.9$).

The catastrophic cooling event, shown in the lower half of Fig.~\ref{fig:10_11_traj_cool} (panels c,d; see also Fig.~\ref{fig:9_traj}d),\footnote{A 3D movie focusing on this event can be found via the link: \url{https://youtu.be/zWn4Co1mLJU}.} is different from the subtle scattering event discussed above. In this case, the particle is accelerated by the ``pick-up'' mechanism to energy $\gamma\gtrsim\sigma$, while moving along a slingshot-like trajectory. Before $t\approx 3.5$, although the velocity of the particle is perpendicular to the local magnetic field (as indicated in Fig.~\ref{fig:10_11_traj_cool}c, and Fig.~\ref{fig:9_traj}c, and d), the effective perpendicular magnetic field, $\tilde{\bm{B}}_\perp$ is almost zero, allowing the particle to avoid being radiatively cooled. At the time $t\approx 3.55$, when the particle enters into the dense flux tube, $\tilde{\bm{B}}_\perp$ increases dramatically (above the critical value, $B_{\rm up}(\gamma_{\rm syn}/\gamma)$), and the particle experience a strong radiative drag force (as shown in the second panel on the lower-right of Fig.~\ref{fig:10_11_traj_cool} with an orange line, $\bm{f}_{\rm drag}$). In this catastrophic cooling episode, the particle loses most of its energy and is further trapped inside the flux tube (also shown in Fig.~\ref{fig:9_traj}e). 

In both the scattering and the catastrophic cooling examples, the full radiation reaction force, given by Eq.~\ref{eqn:cooling}, consists of both synchrotron and curvature contributions. As shown in the top and the bottom panels on the right in Fig.~\ref{fig:10_11_traj_cool}, in both cases the local radius of curvature of the field line\footnote{Calculated as $((\bm{\hat{b}}\cdot \nabla)\bm{\hat{b}})^{-1}$, where $\bm{\hat{b}}\equiv \bm{B}/|\bm{B}|$ is a unit vector along the local magnetic field.} during the deceleration is comparable to the characteristic Larmor radius of the particle with the Lorentz factor $\gamma\approx\sigma$, i.e., it is close to the gyro-radius of the particle. Under these conditions, the synchrotron and curvature radiation mechanisms are indistinguishable, with the radius of curvature of field lines being microscopic and set by the local field perturbations inside the current layer, i.e., the curvature radius is comparable to the characteristic sizes of the flux tubes. 

After the acceleration phase, pairs experience a highly inhomogeneous magnetic field while propagating inside the current sheet. Synchrotron cooling strength for a given particle with a Lorentz factor $\gamma$ is controlled by the effective perpendicular magnetic field strength at the position of the particle, $\tilde{B}_{\perp}$. Assuming a balance between the energy gain from the electric field of characteristic strength, $|e| E_{\rm rec}$, and the radiative drag force, $|e| E_{\rm rec}(\tilde{B}_{\perp}/B_{\rm up})^2(\gamma/\gamma_{\rm syn})^2$, one can define the critical value for the effective magnetic field: $\tilde{B}_{\perp}\approx B_{\rm up}(\gamma_{\rm syn}/\gamma)$. As discussed above, pairs can experience short episodes of stronger $\tilde{B}_\perp$, when they rapidly cool down on the timescales of gyration. This condition thus implies, that in simulations with a strong cooling, the highest-energy pairs (with $\gamma\approx \sigma$) that sustain their motion for long periods of time experience a significantly weaker effective perpendicular magnetic field, either biasing towards local weak-field regions or outflows from the X-points. 

To directly measure the effect of energy-dependent field inhomogeneities, in Fig.~\ref{fig:11_hist} we plot the distribution of energies, $\gamma$, and effective magnetic field strengths experienced by particles, $\tilde{B}_\perp$, for pairs in simulations with varying cooling strength. The dashed line indicates the position of the critical magnetic field, $\tilde{B}_\perp\approx B_{\rm up}(\gamma_{\rm syn}/\gamma)$. Evidently, for all the cases particle distributions do not cross the critical threshold, with the highest-energy particles, $\gamma\approx\sigma$, following trajectories with $\tilde{B}_\perp\lesssim B_{\rm up}(\gamma_{\rm syn}/\gamma)$ \citep[cf.][]{Hakobyan2022b}. Importantly, the combination of particle energy, $\gamma$, and the effective magnetic field it experiences, $\tilde{B}_\perp$, determines the peak energy of synchrotron emission, $\varepsilon_p\propto \gamma^2\tilde{B}_\perp$, and in the context of the highest energy pairs, this determines the cutoff of the overall radiation spectrum. The results shown in Fig.~\ref{fig:11_hist} suggest that the naive prediction of the radiation spectrum from the pair distribution (or vice-versa) typically used in one-zone models can be significantly modified by this non-trivial dependence of $\tilde{B}_\perp$ on the particle energy, especially in systems with dynamically strong synchrotron cooling.

\begin{figure}
\includegraphics[width=\columnwidth]{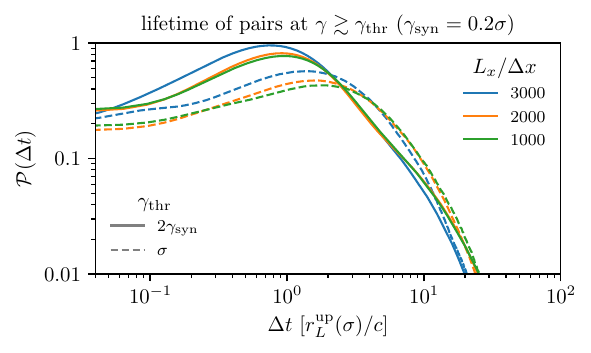}
\caption{Distribution of the lifetimes of strongly cooled pairs at high energies ($\gamma \gtrsim 2 \gamma_{\rm syn}$ and $\gamma \gtrsim \sigma$) for different sizes of the simulation domain. For both thresholds, the lifetime is close to one gyro-period of the most energetic particles, $r_L^{\rm up}(\sigma)/c$ and, importantly, is independent of the domain size.}
\label{fig:12_lifetime}
\end{figure}

As we have seen above, in strongly cooled simulations inhomogeneities of the magnetic field in the current layer do not allow pairs to accelerate indefinitely. In particular, we can estimate the characteristic timescale of catastrophic energy losses for the highest-energy pairs as $t_c \lesssim \gamma / (\beta_{\rm rec}\omega_L)\approx \beta_{\rm rec}^{-1}(r_L^{\rm up}(\sigma)/c)(\gamma/\sigma)$, where $\omega_L=|e|B_{\rm up}/(m_e c)$.\footnote{Here we used the definition of $\gamma_{\rm syn}$ from \eqref{eqn:gammaSyn} and approximated $\tilde{B}_\perp\approx B_{\rm up}(\gamma_{\rm syn}/\gamma)$. Then one can write $\gamma/t_c = f_{\rm drag}/(m_e c) = (4\pi m_e c)^{-1}\sigma_T B_{\rm \perp}^2 \gamma^2$, and substitute $\gamma_{\rm syn}$, and $\tilde{B}_\perp$.} We directly measure the continuous residence time, $\Delta t$, at high energies, $\gamma\gtrsim\gamma_{\rm thr}$, for all the pairs in our strongly cooled runs ($\gamma_{\rm syn}=0.2\sigma$). The distribution of residence times, normalized to $r_L^{\rm up}(\sigma)/c$, is shown in Fig.~\ref{fig:12_lifetime} for two values of $\gamma_{\rm thr}$: $2\gamma_{\rm syn}$, and $\sigma$. To ensure our results are independent of the size of the box, we also over-plot the results with varying domain sizes and separation of scales. Evidently, most of the particles spend a very short amount of time, $\Delta t \lesssim t_c$, at high energies, after which they rapidly cool down, as we have discussed in the example above. Notice, that this timescale is typically much shorter than the light-crossing time of our simulation domain, $L_x / c$, which implies two important points. First of all, this indicates that our simulation domain is large enough so that the pairs are free to experience many acceleration-deceleration events during their lifetime inside the current layer. Despite this, pairs are unable to gain energy indefinitely in successive acceleration events, as each of these episodes is followed by a rapid energy loss. Second, the statistical balance between these two episodes essentially enforces the critical field criterion, $\tilde{B}_\perp\gamma \approx B_{\rm up}\gamma_{\rm syn}$, allowing us to extrapolate the upper bound on pair acceleration to the current layers with realistic scale separations.

\section{Radiation from the current sheet}
\label{sec:radiation} 

\begin{figure}
\includegraphics[width=\columnwidth]{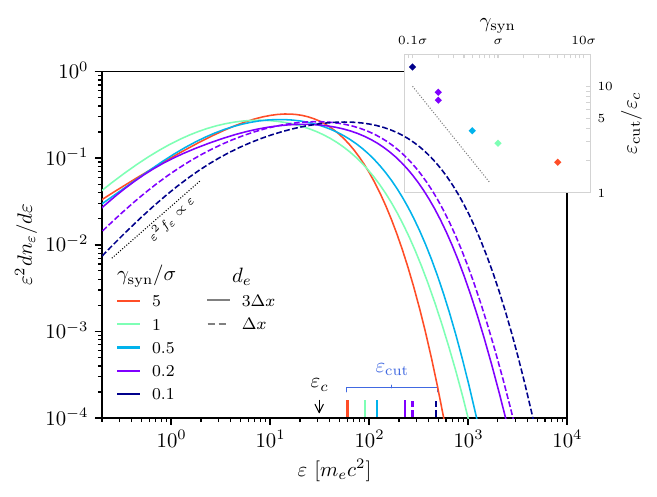}
\caption{Normalized flux density of the emitted radiation for different cooling regimes. The additional panel shows the position of the exponential cutoff of the spectra as a function of the cooling strength. Current sheets with stronger cooling produce a more extended emission spectrum, with significant power above the burnoff limit, $\varepsilon_{\rm c}$, and a cutoff being in good agreement with the prediction of Eq.~\ref{eqn:HighEnergy}: $\varepsilon_{\rm cut}\sim \varepsilon_{\rm c} (\sigma/\gamma_{\rm syn})$. Current sheets with weak cooling, on the other hand, produce almost no emission above the burnoff limit.}
\label{fig:13_spectra}
\end{figure}

The primary probe of the underlying acceleration physics from the reconnecting sheets in astrophysical systems is the properties of the outgoing high-energy radiation. So far we have only focused on the acceleration dynamics of ions and pairs under the influence of the synchrotron drag force, in this section we study the properties of the produced emission.

In Sec.~\ref{subsec:LepAcc}, we demonstrated that in the weak-to-moderate cooling regime, $\gamma_{\rm syn}\gtrsim \sigma$, pair acceleration is limited to $\gamma_{\rm max}\lesssim \gamma_{\rm syn}$. We expect the spectrum of the high-energy radiation to cut off at energies comparable to the burnoff limit, $\varepsilon_c$, from \eqref{eqn:burnoff}. In contrast, in the strong cooling regime, $\gamma_{\rm syn}\lesssim \sigma$, pairs can accelerate to $\gamma_{\rm max}\sim \sigma$ (see, e.g., Fig.~\ref{fig:5_energy_distributions}), and we expect the radiation spectrum to extend above the burnoff limit, as follows from Eq.~\eqref{eqn:photonEn}. Following the discussion in Sec.~\ref{subsec:photons}, the estimate of the cutoff energy for the strong cooling regime can be obtained from the critical value of the effective perpendicular magnetic field, $\tilde{B}_\perp\approx B_{\rm up}(\gamma_{\rm syn}/\gamma)$:
\begin{equation}
  \varepsilon_{\rm cut} = \varepsilon_c \left(\frac{\tilde{B}_\perp}{B_{\rm up}}\right)\left(\frac{\gamma_{\rm max}}{\gamma_{\rm syn}}\right)^2 = \varepsilon_c\left(\frac{\sigma}{\gamma_{\rm syn}}\right).
    \label{eqn:HighEnergy}
\end{equation}

\begin{figure*}
\includegraphics[width=\textwidth]{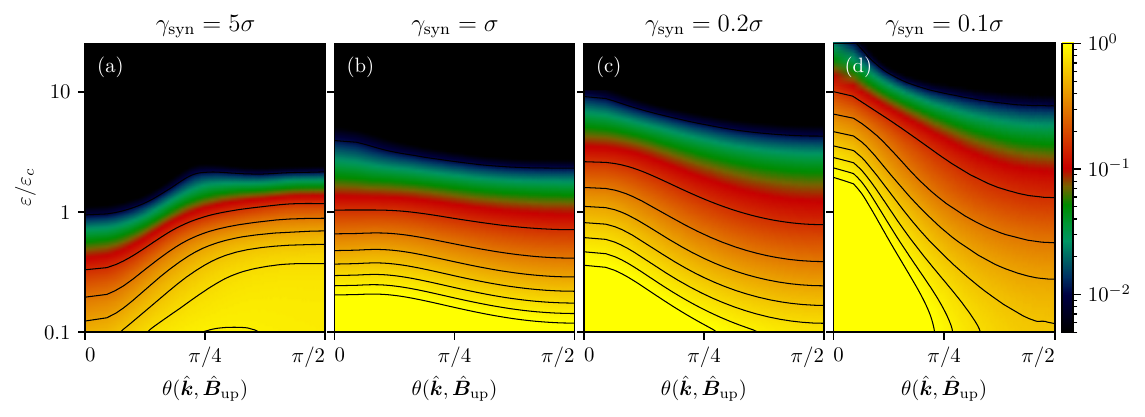}
\caption{Angular anisotropy of the radiation at different photon energies for different cooling regimes. The angle is measured with respect to the direction of the upstream magnetic field $\bm{B}_{\rm up}\equiv B_x\hat{\bm{x}}$. In the weak cooling case, $\gamma_{\rm syn}=5\sigma$, emission above the burnoff limit is significantly suppressed, and most of the high-energy photons are emitted perpendicular to the upstream magnetic field. On the other hand, in the strong cooling regime, most of the high-energy photons above the burnoff limit are emitted along the upstream magnetic field.}
\label{fig:14_beaming}
\end{figure*}

To verify our expectations, in Fig. \ref{fig:13_spectra} we show photon spectra, $\varepsilon^2 dn_\varepsilon/d\varepsilon\equiv \varepsilon^2 f_\varepsilon$, from simulations with different cooling regimes. Each of these spectra is averaged in the interval of $1.5 < tc/L < 4.5$. To construct the spectrum, we assume that every particle emits a continuous synchrotron spectrum \citep[e.g.,][]{1979rpa..book.....R}: 
\begin{equation}
    \varepsilon \frac{dn_\varepsilon}{d\varepsilon} \propto \frac{\varepsilon}{\varepsilon_p}\int_{\varepsilon/\varepsilon_p}^\infty K_{5/3}(x) dx, 
    \label{eqn:syncKernel}
\end{equation}
where $K_{5/3}$ is the modified Bessel function, and the energy $\varepsilon_p$ is defined according to Eq.~\eqref{eqn:photonEn}. All of the spectra rise linearly at low energies, $\varepsilon^2 f_\varepsilon \propto \varepsilon$, which is expected from the hard power-law of reconnection-accelerated particles, $f\sim \gamma^{-1}$. The rise is then followed by a broad peak, and a near-exponential decay at high energies. To find the position of the exponential cutoff, we fit the high-energy parts of the spectra as $\varepsilon^2 f_\varepsilon\propto \exp{(-\varepsilon/\varepsilon_{\rm cut})}$. The inferred values, $\varepsilon_{\rm cut}$, are shown in Fig.~\ref{fig:13_spectra} with colored vertical lines, as well as in the inset panel at the top right. Evidently, our results are consistent both with the prediction of Eq.~\eqref{eqn:HighEnergy} for the strong cooling, $\varepsilon_{\rm cut}\approx \varepsilon_c(\sigma/\gamma_{\rm syn})$, as well as for the weak cooling, where the cutoff approaches $\varepsilon_{\rm cut}\approx \varepsilon_c$. 

The angular distribution of the produced emission is another milestone toward applying our results to realistic astrophysical systems. Previous analysis of the emission anisotropy was done theoretically \citep[e.g.,][]{Uzdensky2011, Cerutti2012b}, in 2D simulations \citep{Cerutti2012a, Cerutti2012b, Cerutti2013, Kagan2016, Mehlhaff2020} or in 3D simulations with a limited separation of scales \citep{Cerutti2014}. In Fig.~\ref{fig:14_beaming} we show the distribution of photon energies, $f(\theta, \varepsilon)$, with respect to $\theta$, the angle between the photon (emitted along the instantaneous motion of the electron/positron) momentum and the upstream magnetic field, $\hat{\bm{x}}$, for different values of the cooling strength. In the weakest cooling case, $\gamma_{\rm syn}=5\sigma$ (Fig.~\ref{fig:14_beaming}a), the highest energy pairs are accelerated in the upstream, along the reconnection electric field, $E_{\rm rec} \approx E_y$. This results in high-energy photons being emitting preferentially perpendicular to the upstream magnetic field (\citealt{Zhang2021}, and Sec.~\ref{subsec:LepAcc}). For the strong cooling (Fig.~\ref{fig:14_beaming}c, d), the emission of photons with energies above the burnoff limit, $\varepsilon\gtrsim \text{few}\cdot \varepsilon_c$, is strongly anisotropic and is beamed towards the direction of the upstream field. The bulk of these photons is emitted by the mechanism described in Sec.~\ref{subsec:LepAcc} and particularly demonstrated in the lower half of Fig. \ref{fig:10_11_traj_cool}: the particle is accelerated by the ``pick-up'' mechanism in the X-point outflow, in the direction of the upstream magnetic field (Fig.~\ref{fig:10_11_traj_cool}c), and later produces the short radiation burst, after colliding face-on with the plasmoid.

\begin{figure*}
\includegraphics[width=\textwidth]{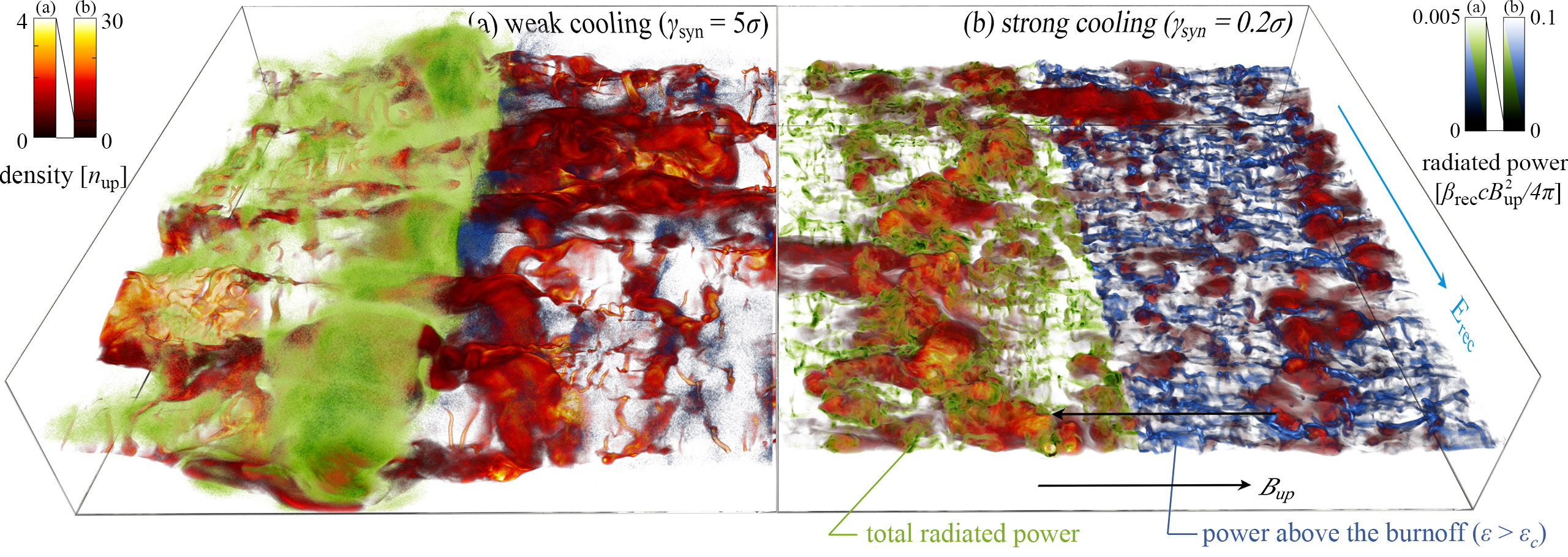}
\caption{Spatial distribution of regions producing the synchrotron radiation shown together with the total plasma density, for the same snapshots as in Fig. \ref{fig:1_3dsnap}. Green volume rendering corresponds to the total synchrotron intensity, while the blue one shows the emission intensity above the burnoff limit, $\varepsilon_c$. Panel (a) corresponds to the weak cooling, while panel (b) is for the strong cooling. The emission in the strongly cooled current sheet is mainly associated with the peaks of the RDKI and the outer shells of the plasmoids (both for the total emission and the emission above the burnoff). The much fainter emission from the weakly cooled current sheet (note the difference in color scales between the two regimes) is significantly more uniform, with the emission above the burnoff being essentially absent.}
\label{fig:15_emissionmap}
\end{figure*}

To understand the spatial distribution of emitting regions, in Fig.~\ref{fig:15_emissionmap} we plot a volumetric map of the radiated synchrotron power, $P_{\rm sync}=\sum_{e^\pm}\bm{F}_{\rm rad}\cdot \bm{v}$, with the $\bm{F}_{\rm rad}$ defined in Eq.~\eqref{eqn:cooling} for simulations with weak (Fig.~\ref{fig:15_emissionmap}a), and strong cooling (Fig.~\ref{fig:15_emissionmap}b), (the summation, $\sum_{e^\pm}$, is done over all the pairs in the given cell). We separately plot the total radiated power (green) and the power carried by the photons above the burnoff limit, $\varepsilon>\varepsilon_c$ (blue). For visual guidance, we use the same snapshot and viewing angle as in Fig.~\ref{fig:1_3dsnap}, and also plot the total plasma density. To properly quantify the radiated power, we normalize it by the incoming Poynting flux, $\beta_{\rm rec}c B_{\rm up}^2/ (4\pi)$. 

In the case of weak cooling (Fig.~\ref{fig:15_emissionmap}a), the density of radiation losses is very weak, i.e., significantly below the Poynting flux inflow (indicated by the colorbar scales for panels (a) and (b)), substantially uniform, and marginally biased towards plasmoids. This is due to the fact that a substantial fraction of the highest energy particles are accelerated by $E_{\rm rec}$ in the upstream of the current sheet, and trapped by plasmoids. On the other hand, for the strong cooling regime (Fig.~\ref{fig:15_emissionmap}b) a large fraction ($\sim 10\%$) of the incoming Poynting flux is being radiated away, with most of the power going into photons comparable to and above the burnoff limit, $\varepsilon_c$. There are two main zones producing the radiation in this case: the outflows from the X-points, and the boundaries of plasmoids. The former is associated with the emission of the X-point-accelerated particles, which interact with the small-scale field inhomogeneities in the outflows of the current sheet (see also Fig.~\ref{fig:10_11_traj_cool}a, b). Radiation near the plasmoid boundaries is associated with catastrophic cooling events when the energetic particles collide face-on with the strong magnetic field surrounding the plasmoids (Fig.~\ref{fig:10_11_traj_cool}c, d). These particles lose energy very intensively at a very short timescale and are ultimately trapped by plasmoids while having very low Lorentz factors, $\gamma \ll \sigma$ (also visible in Fig.~\ref{fig:9_traj}e). Because of this, in the strong cooling regime, plasmoid interiors have almost no contribution to the total radiated power, in contrast to the weak cooling regime. 

Emission above the burnoff limit (blue regions) is almost negligible in the case of weak cooling (Fig.~\ref{fig:15_emissionmap}a), which is consistent with our discussion above, and is also highlighted in the emission spectra in Fig.~\ref{fig:13_spectra}. For the strong cooling (Fig.~\ref{fig:15_emissionmap}b), the power above the burnoff limit (blue) is similar to the total power (green), with the main difference being that the highest-energy emission is suppressed in plasmoid interiors. Thus, in the strong cooling regime, the inner volume of plasmoids only emit soft photons, with the photons above the burnoff being emitted primarily near their exteriors, in the outflows from the X-points, and at the local field inhomogeneities (such as the RDKI ripples, see also Fig.~\ref{fig:10_11_traj_cool}). 

\section{Discussion}
\label{sec:Discussion}
In this paper, we study the 3D dynamics of magnetic reconnection in radiatively cooled pair-dominated plasma with a small mixture of ions. Our study focuses on different regimes of synchrotron cooling, parametrized by the burnoff Lorentz factor, $\gamma_{\rm syn}$, which quantifies the dynamical importance of cooling compared to acceleration by the reconnection electric field. In the weak cooling regime, $\gamma_{\rm syn} \gtrsim \sigma$, particle acceleration is effective both for pairs and ions, both of which form a power-law distribution function with an index $f(\gamma)\appropto\gamma^{-1.7}$ that extends to energies substantially above $\sigma$. Here, the highest energy particles (ions and pairs) are mainly accelerated by the large-scale reconnection electric field upstream of the current sheet. In contrast to that, in the strong cooling regime, $\gamma_{\rm syn} \lesssim \sigma$, the acceleration of pairs is severely limited by the radiative losses, and their maximum Lorentz factor reaches only $\gamma\approx\sigma$. Surprisingly, the acceleration of uncooled ions in this regime becomes more efficient as they form a very hard power-law distribution, with the distribution approaching $\gamma^{-1}$. As we uncovered in Sec.~\ref{subsec:IonAcc}, this hardening of the power law in ion distribution is due to plasmoids being compressed by radiative cooling, which limits their ability to trap the accelerated particles and prevent acceleration in the reconnection electric field. Simultaneously, the strong cooling increases the filling fraction of the X-points in the current layer, effectively enhancing the reconnection rate and, as a result, the strength of the electric field.

The difference in acceleration mechanisms results in a number of important implications for the properties of the high-energy radiation. In the case of weak cooling, the spectral cutoff is close to the synchrotron burnoff limit, $\varepsilon_c\approx 16$ MeV, as the particles accelerated in the upstream have large pitch angles and, thus, their energy cannot substantially exceed $\gamma_{\rm syn}$. Conversely, in the case of strong cooling particles are accelerated up to $\gamma\approx \sigma \gtrsim \gamma_{\rm syn}$ in X-points and relativistic outflows, which results in significant emission beyond the burnoff limit, with the cutoff being close to $\varepsilon_{\rm cut}\approx \varepsilon_c (\sigma/\gamma_{\rm syn})$. In addition, the difference in acceleration mechanisms results in a significant difference in the anisotropy of the high-energy radiation. For strong cooling, the most energetic photons are emitted by particles accelerated by the ``pick-up'' mechanism, which primarily move in the direction of the upstream magnetic field. Thus, the highest energy radiation is emitted in the direction along the upstream magnetic field. For the weak cooling, the most energetic pairs propagate along the reconnection electric field, and thus in this regime the most energetic photons are emitted along $E_{\rm rec}$ and perpendicular to the upstream magnetic field $B_{\rm up}$.

\subsection{Acceleration of ions in pulsar magnetospheres}
\label{subsec:pulsar}

\begin{figure}
\includegraphics[width=\columnwidth]{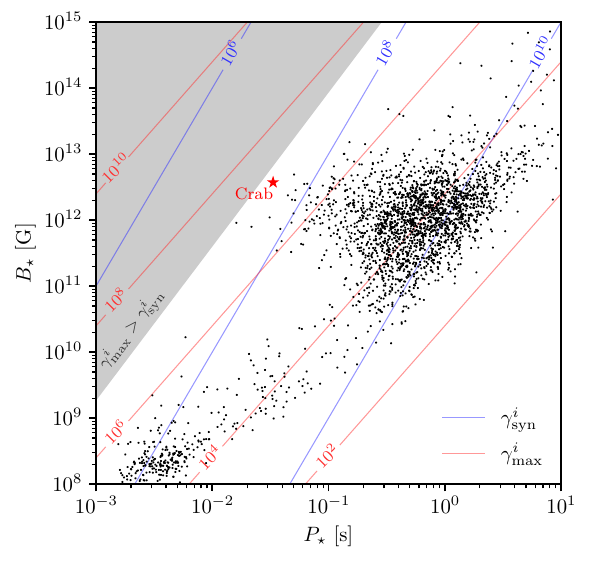}
\caption{Estimates for burnoff limit of protons, $\gamma_{\rm syn}^i$, (blue lines) and the Lorentz factor of protons accelerated in the full potential drop near the light cylinder, $\gamma_{\rm max}^i$, (red lines) for the pulsars in the ATNF catalog. Axes correspond to the rotation period of the star, $P_\star$, and the magnetic field strength near the surface of the star. The shaded region corresponds to the regime, when the synchrotron cooling of protons becomes dynamically important: $\gamma_{\rm max}^i>\gamma_{\rm syn}^i$. Evidently, for all of the observed pulsars, the radiation drag of ions is negligible in the context of ion acceleration.}
\label{fig:pulsars}
\end{figure}

Current sheets in the magnetospheres of pulsars are predominantly filled with the electron-positron plasma, however, a small mixture of ions is also expected \citep{Arons2003, Amato2006, Fang2012, Guepin2020}. Ions can be extracted from the atmosphere of the neutron star near the polar cap region and can carry volumetric and separatrix return current \citep{Timokhin2006, Philippov2018}. Depending on the inclination angle between the magnetic and the rotational axes of the pulsar, a significant fraction of these ions can be carried toward the reconnecting current sheet. Due to their large mass, ions are relatively unaffected by the synchrotron cooling, and may thus be accelerated more efficiently. 

The assumption of negligible cooling for ions can be justified for the observed population of pulsars. Using the definition of the synchrotron burnoff Lorentz factor, Eq.~\eqref{eqn:gammaSyn}, taking the Thomson cross-section of ions, $\sigma_T^i = \sigma_T Z^4/A^2 (m_e/m_p)^2$, $m_i=Am_p$ ($m_p$ is the mass of a proton), and $q_i = Z |e|$, we find that for the ions $\gamma_{\rm syn}^i = A Z^{-3/2}(m_p/m_e)\gamma_{\rm syn}$, with $\gamma_{\rm syn}$ being the burnoff Lorentz factor for pairs (see Sec.~\ref{subsec:photons}). Since most of the high-energy emission takes place in regions of the current layer close to the light cylinder, we may further employ $B_{\rm LC}\approx B_\star (R_\star/R_{\rm LC})^3$ as a proxy for $B_{\rm up}$, with $R_\star$, and $B_\star$ being the radius of the neutron star, and the magnetic field strength near its surface, calculated as $B_\star = 10^{12}~{\rm G} \sqrt{(P_\star/1~{\rm s}) (\dot{P}_\star/10^{-15})}$, and $R_{\rm LC}=cP_\star/2\pi$ being the light cylinder radius ($P_\star$ is the rotation period of the star). We then find

\begin{equation}
    \gamma_{\rm syn}^i \approx 10^{10}~\frac{A}{Z^{3/2}} \left(\frac{P_\star}{1~\textrm{s}}\right)^{3/2}\left(\frac{B_\star}{10^{12}~\textrm{G}}\right)^{-1/2},
\end{equation}

\noindent which can be compared with the total potential drop in the current layer, $q_i \beta_{\rm rec} B_{\rm LC}R_{\rm LC}$, normalized to $m_i c^2$:

\begin{equation}
    \gamma_{\rm max}^i\approx 4\cdot 10^3~\frac{Z}{A}\left(\frac{P_\star}{1~\textrm{s}}\right)^{-2}\left(\frac{B_\star}{10^{12}~\textrm{G}}\right).
    \label{eq:ion_voltage}
\end{equation}

\noindent Then the assumption of negligible cooling for ions holds when $\gamma^i_{\rm syn}\gtrsim \gamma^i_{\rm max}$. We compare these two quantities for protons ($A=Z=1$) for the population of about 3000 radio pulsars from the ATNF catalog \citep{Manchester2005, ATNF} by overplotting the isocontours on the $P_\star-B_\star$ diagram in Fig.~\ref{fig:pulsars}. The gray-shaded region in the parameter space indicates the strong cooling regime for ions, $\gamma_{\rm syn}^i < \gamma_{\rm max}^i$, and evidently, none of the observed radio pulsar fall into that category. This justifies our assumption made throughout the paper on neglecting the radiation drag for the ions. As seen from Fig.~\ref{fig:pulsars}, a small population of the most energetic pulsars with small-enough rotation period and large surface magnetic fields (e.g., Crab) are able to accelerate protons to Lorentz factors $\gtrsim 10^6$, i.e., above PeV energies. For heavier nuclei with $Z\approx A > 1$, the cooling is slightly more efficient, as $\gamma^i_{\rm syn}/ \gamma^i_{\rm max}\propto Z^{-1/2}$, and may thus inhibit their acceleration even for the most energetic pulsars. Acceleration near the light cylinder is also severely radiation-limited for the parameters of fast new-born magnetars \citep[see, e.g.,][]{Arons2003, Fang2012}. In this case, however, if the reconnection continues to be efficient further from the light cylinder, additional acceleration is possible in the wind, where the magnetic field is weaker and the radiative losses are less severe. 

The mass density fraction of ions in pulsar current sheets is low, $f_i\approx (m_i/m_e)\alpha/\lambda \approx 0.01$ (which is the value we employed in our simulations), where $\lambda=n_{\rm \pm}/n_{\rm GJ}\sim 10^4$ is the pair multiplicity, and $\alpha=n_i/n_{\rm GJ}\sim 0.1$ is the ion density relative to the Goldreich-Julian density \footnote{Low fractions of the extracted ions are due to the fact that the bulk of the return magnetospheric current is carried by produced electron-positron pairs \citep[e.g.,][]{Timokhin2013, Chen2014, Philippov2018}.}, \mbox{$n_{\rm GJ}= |\bm{\Omega}\cdot {\bm B}|/2\pi c e$}, where $\bm {\Omega}$ is the angular velocity of the pulsar. The energy budget of even this small fraction of ions, $n_i m_i c^2 \langle\gamma_i\rangle$, can become substantial as they accelerate to high energies, which will inevitably lead to steepening in their distribution function. In the parameter regime where synchrotron cooling is strong for pairs ($\gamma_{\rm syn} < \sigma_{\rm LC}$), and weak for ions ($\gamma_{\rm syn}^i\gtrsim \gamma_{\rm max}^i$), the energy of pairs is limited to $\gamma_{\rm max}\approx \sigma_{\rm LC}$, while the distribution of ions can extend to the full potential drop, $\gamma^i_{\rm max}$, Eq.~\eqref{eq:ion_voltage}. Assuming a power-law index of $-1$ for ions and $\langle \gamma_\pm \rangle \approx \gamma_{\rm syn}$ for pairs (cf. our simulation \texttt{3dx3kCool02}), we may rewrite the total energy budget for both species as:

\begin{equation}
\begin{array}{lcl}
    \mathcal{E}_\pm &\approx& \lambda n_{\rm GJ} m_e c^2 \gamma_{\rm syn},\\
    \mathcal{E}_i &\approx& \alpha n_{\rm GJ} m_i c^2 \gamma^i_{\rm max}.
\end{array}
\end{equation}

\noindent On the other hand, the magnetization near the light cylinder, $\sigma_{\rm LC}$, can be expressed as $\sigma_{\rm LC}= (P_\star/4\pi\lambda) (|e|B_{\rm LC}/m_e c) = 5 \gamma^\pm_{\rm max}/\lambda$, where $\gamma_{\rm max}^\pm = \beta_{\rm rec} |e| B_{\rm LC} R_{\rm LC}/m_e c^2$ is defined as the Lorentz factor of an electron or a positron, which taps $\beta_{\rm rec}$ fraction of the total voltage drop. We can thus find $\mathcal{E}_\pm/\mathcal{E}_i = (5/\alpha)(\gamma_{\rm syn}/\sigma_{\rm LC})$, where we used the relation $m_e \gamma^\pm_{\rm max} = m_i \gamma^i_{\rm max}$. For a Crab-like pulsar with $\gamma_{\rm syn}/\sigma_{\rm LC}\approx 0.1$, we find that pairs dominate in terms of the total energy budget, i.e., $\mathcal{E}_\pm\gg \mathcal{E}_i$, even when the ions are accelerated to the full voltage. This implies that the feedback of ions on the reconnection dynamics is negligible, and the ion spectrum extends to $\gamma_{\rm max}^i$ without significant deviations from the $f(\gamma)\propto\gamma^{-1}$ predicted in our simulations. In fact, we do not observe the spectrum to steepen even when $\mathcal{E}_\pm/\mathcal{E}_i \approx 1/3$, as happens in the simulation \texttt{3dx3kCool02} at late times. 

Another channel of ion acceleration in pulsar magnetospheres is the energy gain by the parallel electric field in the discharge region close to the polar cap. The amplitude of the unscreened voltage is limited by pair cascade, which in young pulsars is triggered when the energetic pairs reach Lorentz factors, $\gamma_{\rm th}\sim 10^6...10^7$. The same potential drop will correspond to the maximum Lorentz factor of protons $\gamma_{\rm th}^i \sim 10^3...10^4$ (smaller by a factor of $m_i / m_e$). These values are much smaller than the ones corresponding to the acceleration in the current sheets of Crab-like pulsars, $\gamma^i_{\rm max} \gtrsim 10^6$, reinforcing the notion that the acceleration in the current sheet beyond the light cylinder is the most efficient channel of ion energization in magnetospheres of young pulsars.  

\subsection{High-energy emission from young pulsars and supermassive black holes}

In Sec.~\ref{sec:radiation}, we discussed the predictions for the observed high energy spectrum from the accelerated pairs. To summarize, in the strong cooling regime applicable in the context of the magnetospheric current sheets of young pulsars, the flux density of the synchrotron emission rises linearly $\nu F_\nu\propto \nu$ (with $\nu$ being the photon frequency) up to the energy comparable to the synchrotron burnoff limit, $16~\text{MeV}$. The broad peak at energies higher than the burnoff is then followed by a steep cutoff at energies close to $16\cdot (\sigma/\gamma_{\rm syn})~\text{MeV}$ \citep[cf.][]{Hakobyan2023}. For the young pulsars, such as the Crab, where the values for the gamma-ray cutoff are between one-to-few GeV \citep{Abdo2013}, our prediction yields $\sigma_{\rm LC} / \gamma_{\rm syn}\sim 100$ (i.e., the cooling regime is strong). For Crab pulsar, where $\gamma_{\rm syn}\approx 4\cdot10^4$ close to the light cylinder, this implies that the magnetization parameter near the current layer is around $\sigma_{\rm LC} \approx 10^6$. Notably, the presence of pairs with energies $\gtrsim \text{few}\cdot 10^6~m_e c^2$ is a strict requirement to explain the pulsed emission above TeV energies in the Crab pulsar \citep{CrabTeV}.

The beaming of the high energy emission is another central outcome of our work. Modeling of gamma-ray lightcurves from global simulations of pulsar magnetospheres \citep{Bai2010,Cerutti2016,Kalapotharakos2018} discovered that in order to explain the observed lightcurves the emission has to be beamed along the magnetic field lines, in the corotating frame of the pulsar. It has thus been not entirely clear how magnetic reconnection in the outer-magnetospheric current sheet could be responsible for such energization, as the naive expectation implied that particles were beamed in the direction of the reconnecting electric field, perpendicular to the direction of the magnetic field. Our findings outlined in Sec.~\ref{sec:radiation} and in Fig.~\ref{fig:14_beaming}, however, indicate that in the regime of strong synchrotron cooling, $\gamma_{\rm syn}<\sigma_{\rm LC}$, most of the high-energy emission is indeed beamed along the upstream magnetic field, as a result of particles being reaccelerated by the ``pick-up'' mechanism, while moving along the relativistic outflows from the X-points in the direction of the upstream field (see also \citealt{Cerutti2012a} for the 2D study).

Magnetospheres of low-luminosity accreting supermassive black holes have also been predicted to host intermittently formed reconnecting current layers with sizes comparable to few-to-ten black-hole gravitational radii \citep[e.g.,][]{Ripperda2020, Ripperda2022}. In particular, for the black hole at the center of the M87 galaxy, due to abundant pair-loading, the regime of reconnection is predicted to be somewhat similar to the reconnection in Crab pulsar \citep{Ripperda2022}, with $\sigma \sim 10^7$, and $\gamma_{\rm syn}\sim 10^6...10^7$. In that regard, our simulations are largely consistent with the results of 2D simulations by \cite{Hakobyan2023b}, where the authors predict that most of the power is radiated at $10...100$ MeV range, with the most energetic pairs, $\gamma\sim \max{(\sigma, \gamma_{\rm syn})} \sim 10^7$, also producing the observed TeV signal (via inverse Compton scattering of soft disk photons). Remarkably, the luminosity ratio between the TeV signal, and the jet power, $0.1\%$, is close to the luminosity ratio between the TeV and GeV signal from the Crab pulsar, suggesting that the Compton amplification parameters for both of these vastly different systems are somewhat similar.

Finally, in most of the discussion above, we focused on the strong cooling regime, where significant emission is expected to be observed above the burnoff limit. One important instance where it is unclear whether this regime applies is the Crab Nebula, where reconnection has been proposed \citep[see, e.g., ][]{Cerutti2014, Lyutikov2016} as a possible mechanism for the gamma-ray flares at energies $\gtrsim 160$ MeV observed by the Fermi satellite \citep{Tavani2011, Buehler2012}. Here, the upstream pair plasma is relativistically hot, $\langle\gamma\rangle\gg 1$, and the characteristic energy gain per particle is given by the hot magnetization parameter, $\sigma_{h}=B^2/(4\pi n m_e c^2 h)$, where $h\sim \langle\gamma\rangle$ corresponds to the plasma enthalpy per particle. The strong cooling regime then applies only if $\sigma_{h} \langle\gamma\rangle \gtrsim \gamma_{\rm syn}\sim 10^9$, where the estimate for $\gamma_{\rm syn}$ is made assuming the strength of the magnetic field is close to milligauss. Our findings suggest that a significant flux at energies above $\gtrsim 160$ MeV can only be produced when $\sigma_h \gtrsim 10\cdot\left(\gamma_{\rm syn}/\langle\gamma\rangle\right)\approx 1000 (\gamma_{\rm syn}/10^9)/(\langle\gamma\rangle/10^7)^{-1}$. It is currently unclear whether such conditions can be realized in the Nebula.\\~\\

\section{acknowledgments}


We thank James Drake, Lorenzo Sironi, and Dmitri Uzdensky for the enlightening conversations, and anonymous referee for useful comments. This work was supported by NSF Grant No. PHY-2231698, and facilitated by Multimessenger Plasma Physics Center (MPPC), NSF Grant No. PHY-2206607. This work was supported by a grant from the Simons Foundation (00001470, AP). Computing resources were provided and supported by the Division of Information Technology at the University of Maryland (\href{http://hpcc.umd.edu}{\texttt{Zaratan} cluster}\footnote{\url{http://hpcc.umd.edu}}). This research is part of the Frontera computing project at the Texas Advanced Computing Center (LRAC-AST21006). Frontera is made possible by NSF award OAC-1818253. The computing resources for this research were partially provided and supported by Princeton Institute for Computational Science and Engineering and Princeton Research Computing. H.H. was partially supported by the U.S. Department of Energy under contract No. DE-AC02-09CH11466.

\bibliography{sample631, refs-hh}{}
\bibliographystyle{aasjournal}

\begin{figure*}[]
\includegraphics[width=\textwidth]{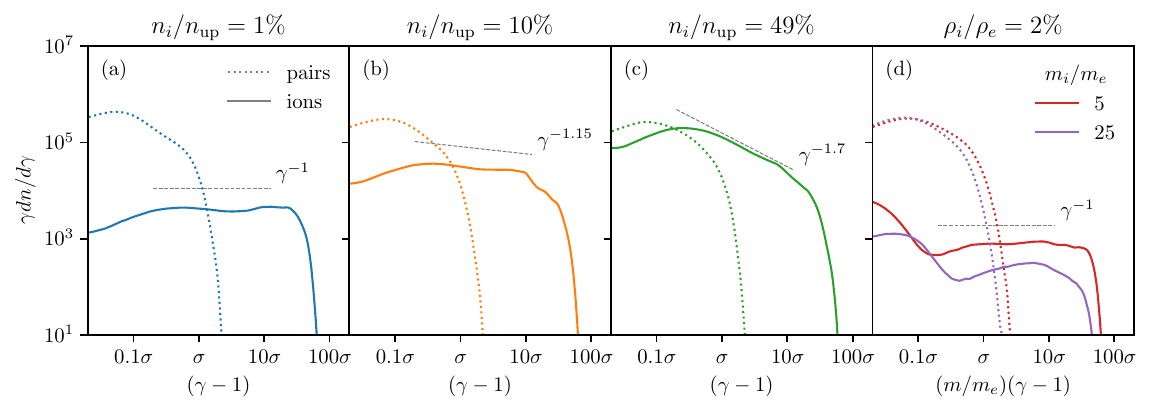}
\caption{Distribution functions of pairs (dashed) and ions (solid) in simulations containing a different fraction of ions, $f_i=n_i/n_{\rm up}=1\%,~10\%,~49\%$ (panels a...c), and different mass ratio $m_i/m_e=5,~25$ (panel d). As long as the mass density fraction of ions is relatively small, $n_i m_i/(n_e m_e) \ll 1$ (which is not the case in panel c), the power-law index of their distribution is insensitive to either $f_i$, or the $m_i/m_e$. The $x$-axis of panel (d) is normalized by the mass of the particles.}
\label{fig:manyIons}
\end{figure*}

\appendix

\section{Varying the fraction of ions and the mass ratio}
\label{app:concentration}

In this appendix, we compare the efficiency of ion acceleration in the current sheet with a varying number density fraction of ions, $f_i=n_i/n_{\rm up}$, as well as the mass ratio, $m_i /m_e$. For this study, we use simulations with a spatial resolution of $d_e=\Delta x$ and a box size $L_x/d_e=1000$. In Sec.~\ref{subsec:IonAcc}, we have demonstrated that in simulations with no cooling and mass ratio $m_i/m_e=1$ the evolution of positrons and ions is identical, and, obviously, there is no difference for different fractions of ions in the simulation. The difference is most pronounced when pairs are strongly cooled. In this section, we compare the evolution of the distribution of ions in the strong cooling regime, $\gamma_{\rm syn}/\sigma=0.2$. 

In Fig.~\ref{fig:manyIons}a, b, c we show a comparison of the distribution of ions measured in the steady state of the reconnection, $t=4 L_x/c$, for varying number density fractions, $f_i=1\%, 10\%, 49\%$, while keeping $m_i/m_e = 1$ (simulation shown in Fig.~\ref{fig:manyIons}a is identical to the previously discussed simulation \texttt{1dx1kCool02}). From these results, we see that as long as the number density contribution of ions is small, $f_i\ll 1$, their distribution remains very hard, $d n_i/d\gamma \propto \gamma^{-1}$ for energies $\gamma\gtrsim\sigma$. In case when the number of positrons is low, and ions constitute almost half of the particle population (Fig.~\ref{fig:manyIons}c), we observe a significant deviation, with the resulting ion distribution, $d n_i/d\gamma \propto \gamma^{-1.7}$, being similar to that observed in the simulations without synchrotron cooling. In this case, the pressure contribution of hot ions to the dynamics of the reconnection layer and plasmoids is no longer negligible. As a result, plasmoids in this case are much larger than in strongly cooled simulations with a lower number of ions, and this leads to more efficient trapping of the accelerated ions and, ultimately, to a steeper power-law index. In all cases, the spectrum of the strongly cooled pairs is not affected, as their acceleration primarily occurs in the X-point region. 

Another limitation of our simulations is the reduced mass ratio, $m_i/m_e$, which we have chosen to be $1$ for most of the discussions throughout the paper. The reason for this choice is to maximize the separation of scales between the Larmor radius of the most energetic ions, $r_L$, and the size of the box, and to minimize the effects of boundary conditions on particle acceleration. To confirm that the reduced mass ratio $m_i/m_e=1$ adequately captures the general properties of the acceleration of ions, we carried out two simulations with the parameters described above, but with larger mass ratios, $m_i/m_e=5$ and $m_i/m_e=25$. To keep the mass fraction of ions constant, $\rho_i/\rho_e = 2\%$, we also decreased their number density fraction, $f_i$. The results of these simulations are shown in Fig.~\ref{fig:manyIons}d. Evidently, the power-law slope of the distribution is only marginally affected by the mass ratio, with the power-law tail extending to energies $\gg m_e \sigma$ (where $\sigma$ takes into account the corresponding mass ratio as shown in Eq.~\eqref{eq:sigma}).

\begin{figure*}[]
\includegraphics[width=\textwidth]{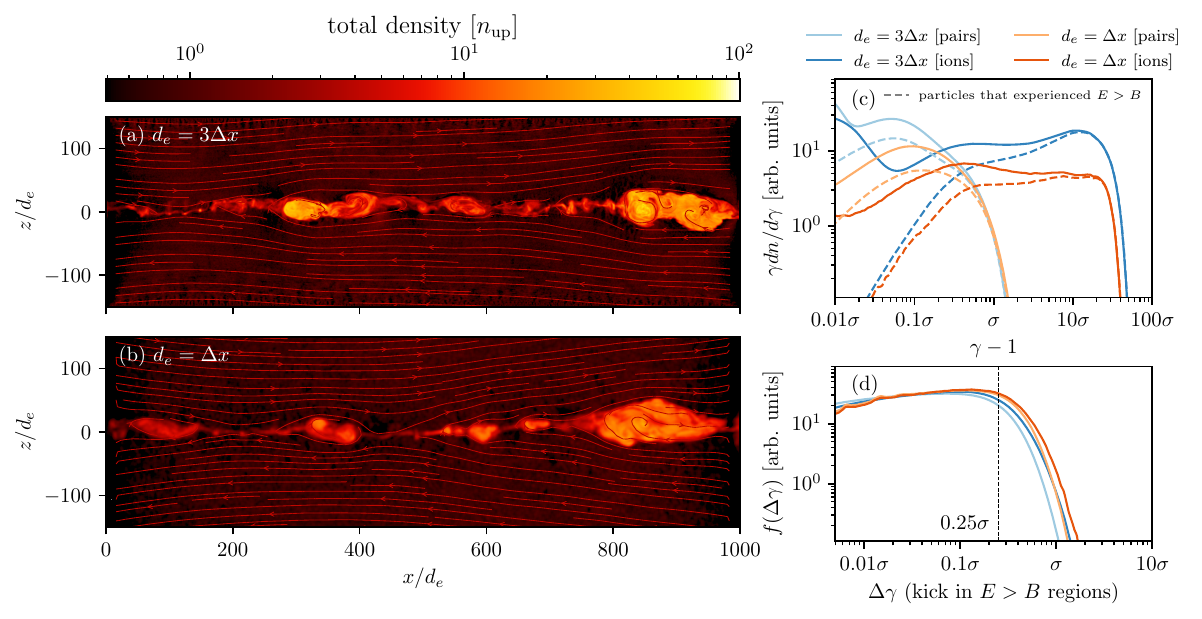}
\caption{Comparison of 2D plasma density slices (a, b) and normalized particle distributions (c, d) for different numerical resolutions $d_e/\Delta x=1,~3$. Dark and bright lines correspond to ions and pairs, while blue and orange correspond to high- and low-resolution runs. The acceleration mechanisms, the statistics of the energy gains in $E>B$ regions, and thus the resulting distributions are largely insensitive to numerical resolution. The compression of plasmoids, on the other hand, is somewhat smaller for the low-resolution run.}
\label{fig:convergence}
\end{figure*}

\section{Varying the resolution}
\label{app:resolution}

To ensure that our results are robust, and do not depend on the resolution, i.e., the number of cells per upstream plasma skin depth, in this section we vary this parameter, $d_e= 1...3 \Delta x$, while keeping $m_i/m_e = 1$, and $f_i=1\%$. For a direct comparison, we will keep other parameters fixed in the simulation: the ratio of the size of the simulation box to the skin depth, $L_x/d_e=1000$, and the magnetization parameter, $\sigma=50$.

We compare two particular cases for the strong cooling, $\gamma_{\rm syn}/\sigma=0.2$ (simulations \texttt{3dx3kCool02} and \texttt{1dx1kCool02}), where the results are expected to be the most sensitive to the resolution, as these simulations demonstrate a strong compression of plasmoids, making the local skin depth potentially difficult to resolve. We choose the efficiency of particle acceleration, i.e., the total distribution function as well as the distribution of energy gains in $E>B$ regions, as the main criteria for comparing different numerical resolutions. The results of this comparison are shown in Fig.~\ref{fig:convergence}. Panels (a) and (b) show slices in $y=0$ plane of the plasma density at time $t=2.7 L_x/c$. Panel (c) shows the normalized distribution of both pairs and ions, as well as the contributions by particles that encountered $E>B$ during their lifetime. Panel (d) shows the normalized distribution functions of the energy gained by both pairs and ions during their flythrough in the $E>B$ region. Evidently, the cross-sections of the current sheet are similar in both cases (Fig.~\ref{fig:convergence}a, b), with the lower-resolution simulation showing less high-density regions. The power-law slopes of the distributions, especially at energies $\gamma\gtrsim \sigma$, are also similar both for the high and the low-resolution runs. The energy gain in the regions of non-ideal field $E>B$ is identical for both simulations, showing that the dynamics around X-points is equally resolved for both simulations.


\end{document}